# A Framework for Transportation and Land Use Integration as a Parallel Constrained Multiple Discrete-Continuous Extreme Value (PC-MDCEV) Home Production Model

Jason Hawkins[a]* and Khandker Nurul Habib[b]

*[a]Department of Civil & Environmental Engineering, University of Nebraska Lincoln, Lincoln, Nebraska, USA; [b]Department of Civil & Mineral Engineering, University of Toronto, Toronto, Canada*

Correspondence: Jason Hawkins Ph.D.
Department of Civil & Environmental Engineering
University of Nebraska Lincoln
900 N 16th St.
Lincoln, NE
68588
Email: jason.hawkins@unl.edu

# A Framework for Transportation and Land Use Integration as a Parallel Constrained Multiple Discrete-Continuous Extreme Value (PC-MDCEV) Home Production Model

Integrated urban models (IUM) typically rely on a measure of accessibility or travel time to form the link between the transportation and land use systems. Such integration does not fully capture the trade-offs made by households in how they spend their limited temporal and monetary budgets. We propose a microeconomic foundation for transportation and land use choice model integration based on the theory of home production. A utility function is developed that considers both household monetary expenditure and individual time use. We address several limitations in previous home production functions. First, the introduction of a parallel constrained multiple discrete-continuous extreme value (MDCEV) structure that allows for the inclusion of multi-person households in the model. Second, travel time is defined as the minimum time required to conduct an activity and deducted from the temporal budget. This assumption has several appealing features. It defines the minimum time to complete an activity as a measure of accessibility. An empirical application is provided for the Greater Toronto Area using a validated synthetic dataset. Empirical results demonstrate an economy of scale in time devoted to home production, analogous to scaling exhibit in market production. It was found that the mix of dwelling types (detached, apartment, etc.) has a significant influence on both time use and consumption. Finally, we provide several directions for future research to advance the practice of urban modelling and better capture the complex dynamics of household decision-making.

## Introduction

Transportation demand model (TDM) development can be traced to the Detroit Metropolitan Area Traffic study (DMATS) in 1953 (Boyce and Williams, 2015). Since this initial effort, TDMs have continued to advance, with the current state of practice being activity-based microsimulation models. It was recognized from the start of TDM development that changes in land use (i.e., built form and location preferences of households and firms) played an essential role in characterizing transportation supply and demand. Among the first efforts to model land use change was an application by Lowry (1964) in Pittsburgh, PA. The integration of the transportation and land use modelling streams can be traced to work by Putman (1970, 1974). Individually, transportation and land use models have seen significant advancements in the methods applied - the adoption of random utility theory and clustered computing - and detailed representation of system components - individual decision-makers rather than traffic flows. However, consistent theoretical integration of transportation and location choice remains an area requiring improvement. In this paper, we propose a conceptual integration framework and provide initial empirical results.

The transportation and land use (built form) systems interact along several dimensions. Dense land use makes higher-order transit (i.e., fixed rail) an effective alternative to single-occupant vehicles through improved service frequency and demand density. Likewise, buildings exhibit supra-linear scaling of water and energy infrastructure with development density, common walls reducing heating requirements, and the ability to employ district heating and cooling (among other benefits) (Newman & Kenworthy, 1999; Zach, Erker, & Stoeglehner, 2019). From an economic perspective, there are production agglomerations associated with urbanization (Duranton and Turner, 2018). Recent advances in technology, as well as the COVID-19 pandemic, have brought into question the role of cities.

We use the *urban amenities* concept (i.e., a variety restaurants and social interactions, physical setting, high quality public services, and physical proximity for social interactions) to motivate our conceptual model (see Hawkins and Habib (2021) for an expanded discussion). Here, the focus is on the combination of monetary benefits (i.e., agglomerations and reduced unit costs) and temporal benefits (i.e., accessibility) as a model link between the monetary and temporal aspects of transportation and land use. The model begins from the premise that household decisions are made subject to temporal and monetary constraints, as argued by Moeckel (2017) and others. Increasingly, transportation demand models account for the time constraint through a scheduling model, often based on Hägerstrand's (1970) time-space prism. The time constraint for a given home-work location choice pair is then captured in the land use model based on travel time skims or accessibility measures derived from TDM outputs. However, this model integration approach does not account for tradeoffs made between consumption of the temporal budget (i.e., time use) and consumption of the monetary budget (i.e., expenditure). The choice of home location is influenced by transportation through the spatial distribution of activities and the travel time to reach these activities. However, it is also influenced by the financial ability of a household to engage in a set of activities after accounting for the purchase price of their home (measured as a monthly mortgage or rent).

This paper's main contribution is a transportation and land use integration framework that begins from microeconomic household decision theory. In the following section, we develop the theoretical foundation for a consistent treatment of transportation and home location decisions, then provide a summary of existing approaches to model integration. The reader should note, while the model underpinnings are drawn from classical economic theory, the motivation is integration between urban land use models (e.g., PECAS (Hunt and Abraham, 2009) and UrbanSim (Waddell, 2002)) with activity-based travel demand models rather than being focused on urban economic models. We then define a home production model subject to temporal and monetary budget constraints, from which we build a conceptual framework for integration of transportation and location decisions. Data for model estimation is an ancillary challenge, which we address through a methodical fusion of existing survey datasets. In the subsequent section, we formulate the model as a household-level likelihood function with individual temporal budgets and a monetary budget that is common across household members. The model is implemented through an extension of the multiple discrete-continuous extreme value (MDCEV) choice model (Bhat, 2005, 2008), with a parallel constrained generalized extreme value (GEV) error structure. An empirical application in the Greater Toronto Area is presented, followed by conclusions and future research directions.

## Theoretical Foundations

### *The Current State of Integration Between Models of Transportation and Land Use Systems*

Many integrated urban models (IUMs) are moving in the direction of microsimulation (Moeckel, 2011; Moeckel et al., 2007; Salvini and Miller, 2005; Waddell et al., 2018). For example, ILUTE is a model system comprising many individual models that follow various theoretical approaches depending on the context (Salvini and Miller, 2005). A standard IUM implementation might use a multinomial logit model for home and work location choice, where TDM travel time skims or mode choice logsum accessibility terms are included in the location choice utility functions. The aggregation of home and work locations across the population then provides an input back into the TDM to generate a new set of network loadings, providing congested travel times for the location choice models in the following time period. Additional details on the evolution of IUM are provided in Hawkins and Habib (2018).

The theory underlying this approach to microsimulation is rarely made explicit but generally traces to principles from systems theory (Smailes, 1971). Within systems theory, the macro-behaviours of urban systems are deemed too complex to represent mathematically. That is, rather than representing residential dynamics as a system of equations, it is simpler to decompose the system into its constituent parts - individuals, households, and firms. However, these lower-level systems may themselves be too complex to represent with any certainty, and the modeler, therefore, represents their parts - births, marriages, relocations, travels - through a sequential simulation. Systems theorists then posit that the macro-behaviour of the system is an emergent property of the collective actions of these individual agents. It is convenient to represent a complex system through an indeterminate number of pseudo-independent models. However, it does not fully capture modeler knowledge about the interactions of these decision systems. Microsimulation models often address this shortcoming in their representation of urban systems through the consistent adoption of random utility theory (RUT), which provides for the calculation of consumer surplus measures as feedback mechanisms between model components. While a good starting point, we argue that this approach lacks a consistent connection with the combination of both time and money constraints faced by households.

## ***Model Integration Through Home Production Theory***

### *Conceptual system overview*

Figure 1 is used to frame the model integration discussion. The standard integration method between transportation and land use systems is via a measure of accessibility or travel time between activities. In the home location choice context, both act as measures of the effective time budget given a location choice. However, location choice and land development are also affected by monetary factors, with more accessible locations generally having higher associated land costs. This relationship between travel and land costs has a long history in the urban economics literature (Alonso, 1964; Mills, 1967; Muth, 1969). A less well-studied component of the relationship between transportation and location is the relationship between rents (or mortgage payments) and effective income for the consumption of other goods and services. Such monetary factors will then translate into impacts on the macroeconomy, as well as activity choice in the transportation model. To close the loop in Figure 1, consider the relationship between macroeconomic models and transportation. Both time and money play a role here. Monetary budgets will be influenced by the cost of travel (e.g., gas prices, transit fares, and vehicle purchase and maintenance costs). Time plays a role in the macroeconomy through agglomeration effects arising in denser urban settings (Rosenthal and Strange, 2008). These relationships bring the discussion to the topic of home production.

[Figure 1 here]

*Home production foundations*

Home production denotes a distinction in the economic literature between goods and services produced in the market and those produced in the home. The conceptual development of the model is attributed to Gary Becker and Jack Mincer in the 1960s (Becker, 1965; Mincer, 1963). Becker began by identifying a lack of consideration of time in the economics literature. He proposed an adjustment of the classical consumption theory to include a time constraint, in addition to the traditional income constraint. Home production theory is best summarized through a series of exemplar comparisons. Consider a household consisting of two adults and two children. In the first scenario, the household must decide how to care for the children during the day. The first option would be to send the children to daycare, in which case both parents would most likely need to work to pay daycare fees. The second option would be for one parent to stay home and care for the children. This would save on daycare, and other household maintenance, costs but lead to a reduced monetary budget. The trade-off in home production theory is generally between 1) additional engagement in the market economy to increase the household budget (i.e., wages multiplied by work time) and 2) devoting household time to the uncompensated production of a good or service. A second scenario is the production of food for consumption by the household. In general, the time required to produce food decreases with an increasing monetary cost. Therefore, the household must decide whether to produce food from scratch, purchase processed food, or purchase a prepared meal. Home production of food requires uncompensated time, while processed food requires additional monetary outlays - therefore, additional income from compensated work.

*Connections between home production theory and IUMs*

Analogous to Becker, the motivation herein is the disproportionate focus of IUMs on time over money. There is a growing literature in urban economics regarding the role of amenities on the attractiveness of a community. Foundational work by Roback (1982) examines the interactions between wages, rents, and urban amenities. Glaeser et al. (2001) challenge a premise that still rings true - whether cities are indeed good for production and bad for consumption. Murphy (2018) takes a different perspective on urban density based on the Roback model. He argues there is increasing productivity as a function of shared land and factors of production with increasing density. For example, an apartment building might have shared laundry facilities, which will reduce the per-household cost of each machine relative to the individual ownership typical of low-density communities. Murphy also draws inspiration from the work of Becker and defines a home production function to examine the effects of density on consumption. He finds evidence for substitution towards market consumption in dense communities. Rappaport (2008, 2009) studies the question of density and urban amenities in several papers. He finds that amenities cause about a fifth of the variation in MSA population density in the United States.

As transportation models increasingly focus on activity scheduling over trip generation (Tajaddini et al., 2020) and land use models become more sophisticated in their treatment of behaviour (Hawkins and Habib, 2018), it becomes possible to introduce a broader treatment of households into IUMs that accounts for both temporal and monetary dimensions.

*Existing work in the transportation literature*

Several authors have developed conceptual models integrating transportation and land use decisions. Jara-Diaz and Martinez (1999) and Eliasson (2010) develop theoretical models that jointly consider location and travel choices. They make several theoretical contributions, which inspired elements of our model. The Eliason model seeks to capture the choice of residential zone, dwelling type, car ownership, trip frequency, and mode choice in an integrated model that includes travel demand. Long-term home location decisions are viewed as being influenced by activity patterns, with the temporal and monetary budgets being reduced with the addition of an activity. Activity patterns then form the basis for trip and mode choice decisions in the short-term transportation demand model. The resulting integrated model is denoted as TILT (Tool for Integrated analysis of Location and Transport). The utility of a location consists of four parts: the indirect utility of income and time net of housing costs and expected total travel times and travel costs, the direct utility of the optimal activity pattern, the direct disutility of the expected travel pattern, and the direct utility derived from location characteristics. The final location choice model is a standard multinomial logit form, with composite terms from each of the utility components, direct and indirect. Application is made to road pricing using a generic city to overcome data availability restrictions.

Arentze, Ettema, and Timmermans (2010) specify a model with utility as a function of the *a priori* attractiveness of the activity, a frequency term, and an exponential decay term for activity engagement. The model includes three activities (in-home, out-of-home maintenance, and out-of-home social) and two generic goods. Integration is made with the PUMA agent-based IUM developed by the same authors (Ettema et al., 2005). Compared with our proposed model, their structure is very simple - five parameters per activity. A simplified application of the model is made for Amsterdam. The budget constraints are largely implicit in the model, and there is no optimization of allocation of time and money between activities. Budgets are considered in the utility function, but a set of decision rules are used in many cases.

More recently, Lopez-Ospina et al. (2015) considered the interaction between short-term daily decisions (e.g., shopping and leisure) and long-term decisions (e.g., home and work location) through a time-hierarchical microeconomic model of activities. Decisions are classified into macro and micro time windows. Short-term decisions are adjusted at a time interval defined by the micro time window. Similarly, long-term decisions are adjusted at a time interval defined by a longer macro time window. The simultaneous adjustment of long-term and short-term decisions is defined as an adjusting point, with long-term decisions influencing short-term decisions subsequent to the adjustment point. A numerical experiment is performed with the model, but no empirical case study is provided by the authors. Martinez (2018) - a coauthor on work by Lopez-Ospina - provides the most recent and closely aligned model structure to that proposed herein. He proposes a general urban system model that starts from an MDCEV model of expenditure choice. In the Martinez model, a short-term decision is made between expenditure on goods and leisure at alternative locations. This expenditure decision is conditional upon a nested logit choice of home and work location. Using a logit (or Fréchet) error distribution provides a straightforward measure of consumer surplus through expected maximum utility. The expectation from the expenditure model can be incorporated into the home location choice as accessibility measures for the consumption of goods and services. However, no empirical application is provided by Martinez (2018). His model would require both the continuous expenditure (or time use) and the location for each activity. Such data is rarely available, so, estimation and prediction would both suffer from data limitations.

*Framing of the proposed model in the literature*

With this literature in hand, we can discuss the proposed model integration. Long-term location decisions are influenced by both accessibility - measured in activity-based transportation models as a space-time constraint - and monetary constraints - measured as a function of income and market conditions. Therefore, it is proposed to treat these decisions within a unified framework that includes time allocation, monetary expenditure, and the constraints on both. Like Eliasson (2010), we frame accessibility as the indirect utility of residual income (net of housing and transportation costs) and residual time (net of travel time) plus the utility of the optimal activity pattern less the disutility of travel. These factors can be captured through a home production function, which can then be fed into a home location choice model. This approach has the benefit of marginal utilities of time, money, and trips being derived from a consistent microeconomic optimization of activity pattern choice. The model introduces new dependencies between land use and transportation models, beyond the single link feedback common in integrated urban models. Activity participation and market expenditures become links between short-term behavior and long-term characteristics of individuals that impact their residential preferences. For example, the decision to socialize with friends at a restaurant after work will reduce the time available to shop for groceries and other goods. If this activity becomes a habit, it will also reduce the money available to perform other activities in the future. The specification of a joint home location-home production model explicitly defines the accessibility that will arise from a particular location choice through its associated activity patterns, residual budgets, and the direct disutility of the expected travel pattern.

## Econometric Model Structure

The description of the econometric model is broken into several components. First, the standard home production model is briefly outlined, and a connection is made to the chosen demand formulation through a multiple discrete-continuous extreme value (MDCEV) model. Then, an extension is made to incorporate home location choice into the model, and a discussion is provided of the role played by time and money in the framework. We then discuss the challenge of multi-member households for home production. Finally, the model structure is developed for the joint time and expenditure choice.

### *Home Production*

We turn now to defining the model of DeSerpa (1971) and its extension by Jara-Diaz (2003), which forms the foundation of our model. DeSerpa places time within the context of a generalized system of consumer demand. The consumer is faced with a two-constraint utility maximization problem. The consumer is subject to the following resource constraints[1]

---

[1] We adopt DeSerpa's notation in this sub-section, which assumes a one-to-one relationship between time use and goods consumption, for consistency and simplicity. We use a slightly altered notation in the development of our model (see Table 1 for a summary).

$$Y = \sum_{i=1}^{n} p_i x_i \tag{1}$$

$$T^o = \sum_{i=1}^{n} t_i \tag{2}$$

where $p_i$ is the unit price, $x_i$ is the consumed quantity, and $t_i$ is the time use for good/activity *i*. Y is the total available monetary budget. $T^o$ is a 24-hour temporal budget. In addition, DeSerpa defines time consumption constraints (representing the minimum time allocation to an activity/good for its consumption) as follows

$$t_i \geq a_i x_i \; \forall i \; \in n \tag{3}$$

where $a_i$ is interpreted as a technologically or institutionally determined minimum time requirement to consume a unit of $x_i$. Technology-constrained activities include movies, reading a book, or eating a meal. Institutional constraints represent limits imposed by institutions, such as speed limits or mandated hours of work. The optimization problem is defined by the following Langrage function

$$L = U(x_1, .., x_n, t_1, \dots, t_n) + \lambda(Y - \sum_{i=1}^{n} p_i x_i) + \mu(T^o - \sum_{i=1}^{n} t_i) - \sum_{i=1}^{n} \kappa_i (t_i - a_i x_i) \tag{4}$$

where $x_n$ and $t_n$ are, respectively, expenditures and time allocated to activity/good *n*, $\lambda$ and $\mu$ are parameters representing the shadow prices on the resource constraints, and $\kappa_i$ on the time consumption constraints.

Our model specification starts from the MDCEV model of Bhat (2005, 2008). The utility function takes the form of a generalized variant of the translated CES utility function

$$U(\mathrm{x}) = \sum_{n=1}^{N} \frac{\gamma_n}{\alpha_n} \psi_n \left[ \left( \frac{x_n}{\gamma_n} + 1 \right)^{\alpha_n} - 1 \right] \tag{5}$$

This takes the form of a linear expenditure system (LES) when $\alpha_k \rightarrow 0 \forall n$

$$U(\mathrm{x}) = \sum_{n=1}^{N} \gamma_n \, \psi_n \ln\left(\frac{x_n}{\gamma_n} + 1\right) \tag{6}$$

In the MDCEV model presented above, $\gamma_n$ is a translation parameter that introduces corner solutions, $\psi_n$ is a baseline marginal utility term (i.e., utility at zero consumption), and $x_n$ is consumed quantity (as in equation 1) and unit prices are set to unity as it is typical to only have consumption available in monetary and not quantity units.

The model is formulated as a direct utility function, which is constrained by a budget constraint on total consumption. This can take the form of a time constraint in the case of time use models or a monetary constraint (as a function of income) in the case of consumption models.

***Location Choice and Home Production Integration***

The time use allocation and expenditure model developed in this research is the first step towards an integrated treatment of home location choice and patterns of activity engagement. The standard home production function assumes fixed home and work locations. However, this assumption may not hold over time, and changes in either location will affect the model. The magnitudes of the constraints on time use and expenditure (i.e., 24 hours less travel time and total household income) are principally governed by the home and work location choices of the household. A household with a work location that provides a higher wage will be able to locate in higher priced - more desirable - areas and consume more goods and services. From the perspective of the employer, on an inter-regional scale, the attractiveness of a city will raise home prices and may require the employer to pay a higher wage rate - higher demand for jobs having a dampening effect on this pattern. On an intra-regional scale, the wage rate is determined mainly by occupation type and exogenous to the model. However, the locations of the workplaces by occupation type will influence the available temporal budget after subtracting transportation time. It may also affect the consumption of out-of-home food, e.g., through proximity to food vendors during the lunch hour. Home location choice affects both time use and consumption (and vice versa for their effect on home location choice). Transportation cost is a component of the transaction cost of an activity or good. The time technology constraints in a standard home production model denote the activity and good qualities (i.e., whether they are deemed as leisure or discretionary), but they also include implicit transaction costs. These costs will vary depending upon the home location choice of the household.

The joint location-time-expenditure decision process comprises a combination of discrete and continuous choices. This feature can be captured in an econometric model through manipulation of the error structure. In the same way that the multinomial logit model can be extended to a nested logit model through assumptions about the underlying error distribution, the MDCEV model can be extended to include structure to the decision process. Pinjari and Sivaraman (2013) consider a similar problem to that considered in the current work – discrete mode choice, conditional upon vacation destination choice and duration. The utility for the joint choice of home location, expenditure, and time use can be represented as follows

$$\max\left(U_q(\mathbf{x}_{ql}, \mathbf{t}_{ql}, \mathbf{t}_{qlw})\right) = \sum_{l=1}^{L}\sum_{k=1}^{K} u_k\left(x_{qlk}\right) + \sum_{l=1}^{L}\sum_{n=1}^{N} \tilde{u}_n\left(t_{qln}\right) + \sum_{l=1}^{L} \tilde{u}_w\left(t_{qlw}\right) \quad (7)$$

where $U_q(\mathbf{x}_{ql}, \mathbf{t}_{ql}, \mathbf{t}_{qlw})$ is a measure of total direct utility from expenditure and time use given a home location $l$. The components of utility are $u_k$ for monetary consumption of $k$ goods, $\tilde{u}_n$ for the time allocated towards $n$ non-work activities, and $\tilde{u}_w$ for the time allocated to the work activity. There are L locations considered by each household and each individual consumes K goods and spends time on N non-work activities. Following the MDCEV structure, the baseline marginal utilities are defined by

$$\psi_{qkl} = exp\left(\boldsymbol{\beta}'_q \boldsymbol{z}_{qlk} + \boldsymbol{\delta}'_q \boldsymbol{x}_{ql} + \epsilon_{qlk}\right) \tag{8.1}$$
$$\psi_{qnl} = \exp\left(\widetilde{\boldsymbol{\beta}}'_q \tilde{\boldsymbol{z}}_{qln} + \widetilde{\boldsymbol{\delta}}'_q \widetilde{\boldsymbol{x}}_{ql} + \tilde{\epsilon}_{qln}\right) \tag{8.2}$$
$$\psi_{qwl} = \exp\left(\widetilde{\boldsymbol{\beta}}'_q \tilde{\boldsymbol{z}}_{qlw} + \widetilde{\boldsymbol{\delta}}'_q \widetilde{\boldsymbol{x}}_{ql} + \tilde{\epsilon}_{qlw}\right) \tag{8.3}$$

where $z_{qlk}$, $\tilde{z}_{qln}$, and $\tilde{z}_{qlw}$ are factors contributing towards the consumption and time use decisions for individual $q$ and $x_{ql}$ and $\tilde{x}_{ql}$ are factors contributing towards the location decision that are specific to individual $q$ but independent of the consumption of a given good or time use activity. Finally, $\epsilon_{qlk}$, $\tilde{\epsilon}_{qln}$, and $\tilde{\epsilon}_{qlw}$ are type I extreme value random shocks accounting for unobserved factors. The conceptual model structure is illustrated in Figure 2 below.

[Figure 2 here]

As in van Nostrand et al. (2013), it is assumed that satiation parameters do not vary across the discrete alternatives. This is logical since one would assume that satiation is a function of the decision-maker and independent of their location choice. A larger γ suggests a stronger preference (or lower satiation) for a good, which will be specific to the individual and not their home location choice. The joint cumulative distribution $F$ of the error terms (dropping $q$s and ignoring heteroskedastic errors across the $Q$ individuals for simplicity of exposition) is $\left(\epsilon_1^*, (\epsilon_{12}, \dots \epsilon_{1k}), (\epsilon_{l2}, \dots \epsilon_{lQ}), \dots (\epsilon_{L1}, \dots \epsilon_{LQ})\right)$. The nested extreme value distributed error term structure has the following cumulative distribution

$$F\left(\epsilon_1^*, (\epsilon_{12}, \dots \epsilon_{1k}), (\epsilon_{l2}, \dots \epsilon_{lK}), \dots (\epsilon_{L1}, \dots \epsilon_{LK})\right) = \left[\exp\left(-\exp\left(\frac{-\epsilon_1^*}{\sigma}\right)\right)\right] \prod_{l=1}^{L} \left[\exp - \left(\sum_{k=1}^{K} \exp\left(\frac{-\epsilon_{lk}}{\sigma\theta}\right)\right)^{\theta}\right] \tag{9}$$

In the above formulation, the error terms for all time use and consumption alternatives are grouped into a nest. A scale parameter, given by $\Theta$, is introduced to capture correlations among the random utility contributions of all goods-home $lk$, non-work time-home $ln$, work time-home $lw$ alternatives sharing a home location choice.

***The Role of Time and Money Thresholds***

Minimum time and money thresholds define the distinction between leisure and mandatory activities. Leisure activities are those activities to which an individual devotes more than the minimum time. In empirical applications, these thresholds have been exogenously defined for both time and expenditure (Astroza et al., 2017; Hössinger et al., 2019; Jara-Díaz et al., 2008; Jokubauskaitė et al., 2019). In addition, it is the standard practice to specify the set of leisure activities exogenously and uniformly for all individuals. For example, Astroza et al. (2017) define minimum time constraints as the minimum observation value among respondents in the time use and expenditure surveys used to construct their input dataset. Hamermesh (2019) illustrates the arbitrary nature of such assumptions through the example of his morning run. One might consider this activity leisure, but Hamermesh notes that his run also provides an opportunity to clear his mind and think about his research. More generally, the distinction between work and leisure is primarily based on a philosophical argument about values. As noted by Joshua Greene (2014) in his book Moral Tribes, the utilitarian definition of satisfaction should include anything which measurably improves the state of society. This might include intellectual pursuits of knowledge, truth, or education; civic pursuits of justice and freedom; or personal character pursuits of honesty and creativity. Any of these pursuits can be found in work activities, and therefore ascribed a positive utility, but they are not uniformly valued by all individuals, nor uniformly associated with the work activity.

Our treatment of time and expenditure differs from the above. Rather than using a fixed minimum time across individuals, we rely on the translation term, $\gamma$, in the MDCEV model as a measure of satiation. This specification provides a continuous measure of relative satiation – larger translation parameters denote activities that individuals are willing to spend time on. Similar to existing work in the area (Astroza et al., 2017; Eliasson, 2010; Jara-Díaz and Martínez, 1999) we subtract travel time from the total daily budget. The first advantage of this approach is that it speaks to the concern of Jokubauskaite et al. (2019) that the positive utility of consumption assumption implied by the MDCEV model does not hold for travel time, which is a derived demand with a negative utility. More importantly, it has the feature of integrating an accessibility measure into the home production model. Households that choose less accessible home locations will have a technological (i.e., distance and travel speed) constraint placed on their out-of-home activity participation via a reduced daily time budget. This treatment of time means that the joint model with home location would consider the accessibility effect of alternate home locations on activity participation and time allocation. Further, no minimums are placed on expenditures. As in the case of other empirical home production applications, these deviations from the theoretical model definition introduce caveats to the interpretation of results. First, all activities in the model satisfy the leisure definition of exceeding the minimum given by travel time for out-of-home and zero time for in-home activities. Second, the lack of minimum thresholds for expenditures introduces the possibility of very small totals for some categories. However, unlike time, it is plausible that a person could spend only a few dollars on certain activities without violating a reasonable minimum. For example, a person might choose to purchase an ice cream cone once per year as their only out-of-home food expenditure.

***The Role of Time and Money Budgets***

While removing travel time from the time budget addresses the negative utility concern associated with its application in MDCEV, it makes the time budget endogenous to the location choice. That is, the time budget will depend on the travel time implied by a particular home location choice. Pinjari et al. (2016), note that almost all applications of MDCEV in the literature assume that the total budget is fixed for each individual. They propose a stochastic frontier approach to estimate the maximum time budget an individual is willing to incur. Such a stochastic frontier could be estimated conditional upon home location choice to obtain a time budget measure that could be forecast across location alternatives.

In the same way that the time budget is endogenous to the home location choice, the monetary budget also exhibits this effect. A mortgage payment (or rent) will reduce the available monetary budget as a function of locational price variations. This cost could be subtracted from the individual monetary budget. However, in the empirical case study, we keep dwelling cost in the model as it does not suffer from the same negative utility concern as travel time. The UrbanSim land use model by Waddell (2002) provides an operational IUM example of a similar approach. UrbanSim includes a hedonic price model, which provides prices that feed into the utility for a home location choice model.

Taken together, the endogenous treatment of time and money provides an integrated framework for travel and location choice. Alternative location choices will have heterogenous temporal budgets as a function of travel time. Similarly, the monetary budget available for day-to-day expenses will be conditional upon dwelling costs.

### *The Role of Household Structure*

The second challenge is the extension of the model to multi-person households. Astroza et al. (2017) make the simplifying assumption of only considering single-person households. However, for the model to be of operational use, it must accommodate multi-person households. We propose a parallel constrained individual utility function as a first step towards addressing this issue. In our framework, individuals have temporal budgets that will depend upon the activities they choose to conduct and the required travel time between activity locations. These individual time use utilities are then components of a utility function with a common constraint on monetary expenditure across individuals within a household. The members of a household are assumed to share a monetary budget, which is apportioned among members. There is a rich literature on intrahousehold budget allocation. Kooreman and Wunderink (1996) classify wage allocation as one of the "whole wage": typically managed by a non-working spouse, "allowance": working members dole out money, "pooling": all household members have access to all resources, or "independent management": each worker has access to only their wages. Chiappori and Mazzocco (2017) provide further insights into the long-term dynamics of household budget allocation. They define two resource allocation phases. In the first phase, the household allocates a fixed budget given a static state and short-term time horizon. In the second phase, the household allocates a dynamic budget across an intertemporal set of expenditures over its lifetime. Following these phases, a third phase is defined wherein the household budget is allocated among its members.

### *An MDCEV Home Production Model*

With these barriers addressed, the model can be outlined in detail. An optimization function can be defined for individuals $q$ ($q$ = 1,2,....Q), consuming goods $x_k$ ($k$=1,2,...K), allocating time $t_n$ to non-work activities ($n$=1,2,...N), and allocating time $t_w$ to the work activity. It takes the form

$$\max\left(U_q(\mathbf{x}_q, \mathbf{t}_q, \mathbf{t}_{qw})\right) = \sum_{k=1}^{K} u_k\left(x_{qk}\right) + \sum_{n=1}^{N} \tilde{u}_n\left(t_{nq}\right) + \tilde{u}_w(t_{wq}) \tag{10}$$

subject to the following constraints

$$\sum_{k=1}^{K} p_{qk}\, x_{qk} = E_q + \omega_q t_{qw} \tag{11.1}$$

$$\sum_{n=1}^{N} t_{qn} + t_{qw} = T_q \tag{11.2}$$

where $E_q$ is a monetary endowment beyond hourly wages (e.g., retirement savings, gifts, etc.) and $T_q$ is total time available (i.e., 24-hours per day). Work time ($t_{qw}$) appears in both the temporal and monetary budget constraints, which introduces an additional level of complexity to the model formulation compared with the standard MDCEV model. This difference introduces a multidimensional constraint. Fortunately, this complication can be isolated to the determination of $t_w$, and the other decision variables (i.e., $t_{qn}$ and $x_{qk}$) can be determined in the usual way. The following are the utility function definitions used in the model

$$u_k(x_{qk}) = \begin{cases} \psi_{qk} x_{qk} & \text{if } x_{qk} = 0 \\ \gamma_k \psi_k \ln\left(\dfrac{x_{qk}}{\gamma_k} + 1\right) & \text{if } x_{qk} > 0 \end{cases} \tag{12.1}$$

$$\tilde{u}_n(t_{qn}) = \begin{cases} \tilde{\psi}_{qn} t_{qn} & \text{if } t_{qn} = 0 \\ \tilde{\gamma}_n \tilde{\psi}_n \ln\left(\dfrac{t_{qn}}{\tilde{\gamma}_n} + 1\right) & \text{if } t_{qn} > 0 \end{cases} \tag{12.2}$$

$$\tilde{u}_w(t_{qw}) = \begin{cases} \tilde{\psi}_{qw} t_{qw} & \text{if } t_{qw} = 0 \\ \tilde{\gamma}_w \tilde{\psi}_w \ln\left(\dfrac{t_{qw}}{\tilde{\gamma}_w} + 1\right) & \text{if } t_{qw} > 0 \end{cases} \tag{12.3}$$

The Lagrangian function for individual $q$ is defined by

$$\mathcal{L}_q = U_q(\mathbf{x}_q, \mathbf{t}_q, \mathbf{t}_{qw}) + \lambda_q\left(E_q + \omega_q t_{qw} - \sum_{k=1}^{K} p_{qk}\, x_{qk}\right) + \mu_q\left(T_q - t_{qw} - \sum_{n=1}^{N} t_{qn}\right) \qquad (13)$$

where $\lambda_q$ and $\mu_q$ are Lagrangian multipliers for the temporal and monetary constraints, representing the marginal utilities of time and consumption. The KKT first-order conditions are defined by

$$u_k'(x_{qk}^*) - \lambda_q p_{qk} = 0 \quad \text{if } x_{qk}^* > 0, k = 1,2, \dots K \qquad (14.1)$$
$$u_k'(x_{qk}^*) - \lambda_q p_{qk} < 0 \quad \text{if } x_{qk}^* = 0, k = 1,2, \dots K \qquad (14.2)$$
$$\tilde{u}_n'(t_{qn}^*) - \mu_q = 0 \quad \text{if } t_{qn}^* > 0, n = 1,2, \dots N \qquad (14.3)$$
$$\tilde{u}_n'(t_{qn}^*) - \mu_q < 0 \quad \text{if } t_{qn}^* = 0, n = 1,2, \dots N \qquad (14.4)$$
$$\tilde{u}_w'(t_{qw}^*) + \omega_q\lambda_q - \mu_q = 0 \quad \text{if } t_{qw}^* > 0, w = 1,2, \dots W \qquad (14.5)$$
$$\tilde{u}_w'(t_{qw}^*) + \omega_q\lambda_q - \mu_q < 0 \quad \text{if } t_{qw}^* = 0, w = 1,2, \dots W \qquad (14.6)$$

where $u_{qk}'$, $\tilde{u}_{qn}'$, and $\tilde{u}_{qw}'$ are the marginal utility functions defined by

$$u_k'(x_{qk}^*) = \begin{cases} \psi_{qk} & \text{if } x_{qk}^* = 0 \\ \psi_{qk}\left(\dfrac{x_{qk}^*}{\gamma_k} + 1\right)^{-1} & \text{if } x_{qk}^* > 0 \end{cases} \qquad (15.1)$$

$$\tilde{u}_n'(t_{qn}^*) = \begin{cases} \tilde{\psi}_{qn} & \text{if } t_{qn}^* = 0 \\ \tilde{\psi}_{qn}\left(\dfrac{t_{qn}^*}{\tilde{\gamma}_n} + 1\right)^{-1} & \text{if } t_{qn}^* > 0 \end{cases} \qquad (15.2)$$

$$\tilde{u}_w'(t_{qw}^*) = \begin{cases} \tilde{\psi}_{qw} & \text{if } t_{qw}^* = 0 \\ \tilde{\psi}_{qw}\left(\dfrac{t_{qw}^*}{\tilde{\gamma}_w} + 1\right)^{-1} & \text{if } t_{qw}^* > 0 \end{cases} \qquad (15.3)$$

The optimal consumption (of goods) and time allocation (to activities) satisfy the KKT conditions in equations 14.1-14.6 and the constraints in equations 15.1-15.3. The individual must consume at least one good and participate in at least one non-work activity, represented by a Hicksian composite good (see (Bhat, 2008) for details). The disallowance of corner solutions for the outside good and normalization of its satiation parameter to zero simplifies the conditions as follows

$$\psi_{q1}(x_{q1}^*)^{-1} - \lambda_q p_{q1} = 0 \qquad (16.1)$$

$$\tilde{\psi}_{q1}\left(t_{q1}^{*}\right)^{-1} - \mu_q = 0 \quad (16.2)$$

$$\tilde{\psi}_{qw}\left(\frac{t_{qw}^{*}}{\gamma_w}\right)^{-1} + \lambda_q\omega_q - \mu_q = 0 \quad (16.3)$$

From the conditions in equations 16.1 and 16.2, $\lambda_q$ and $\mu_q$ are given by

$$\lambda_q = \frac{\psi_{q1}}{p_{q1}}\left(x_{q1}^{*}\right)^{-1} = \frac{u_k'\left(x_{q1}^{*}\right)}{p_{q1}} \quad (17.1)$$

$$\mu_q = \tilde{\psi}_{q1}\left(t_{q1}^{*}\right)^{-1} = \tilde{u}_n'\left(t_{q1}^{*}\right) \quad (17.2)$$

Similar conditions to Astroza et al. (2017) can be derived by substituting equations 17.1 and 17.2 into equations 14.1-14.6. For non-workers, this gives

$$\frac{u_k'\left(x_{qk}^{*}\right)}{p_{qk}} = \frac{u_k'\left(x_{q1}^{*}\right)}{p_{q1}} \quad \text{if } x_{qk}^{*} > 0, k = 1,2, \dots K \quad (18.1)$$

$$\frac{u_k'\left(x_{qk}^{*}\right)}{p_{qk}} < \frac{u_k'\left(x_{q1}^{*}\right)}{p_{q1}} \quad \text{if } x_{qk}^{*} = 0, k = 1,2, \dots K \quad (18.2)$$

$$\tilde{u}_n'\left(t_{qn}^{*}\right) = \tilde{u}_n'\left(t_{q1}^{*}\right) \quad \text{if } t_{qn}^{*} > 0, n = 1,2, \dots N \quad (18.3)$$

$$\tilde{u}_n'\left(t_{qn}^{*}\right) < \tilde{u}_n'\left(t_{q1}^{*}\right) \quad \text{if } t_{qn}^{*} = 0, n = 1,2, \dots N \quad (18.4)$$

and for workers, it gives the additional conditions

$$\tilde{u}_w'\left(t_{qw}^{*}\right) = \tilde{u}_n'\left(t_{q1}^{*}\right) - \omega_q\frac{u_k'\left(x_{q1}^{*}\right)}{p_{q1}} \quad \text{if } t_{qw}^{*} > 0, w = 1,2, \dots W \quad (18.5)$$

$$\tilde{u}_w'\left(t_{qw}^{*}\right) < \tilde{u}_n'\left(t_{q1}^{*}\right) - \omega_q\frac{u_k'\left(x_{q1}^{*}\right)}{p_{q1}} \quad \text{if } t_{qw}^{*} = 0, w = 1,2, \dots W \quad (18.6)$$

Stochasticity and decision-maker variation can be introduced to the model through parameterization of the baseline marginal utility and additive IID error terms as follows

$$\psi_{qk} = exp\left(\boldsymbol{\beta}_{qk}'\boldsymbol{z}_{qk} + \epsilon_{qk}\right) \quad (19.1)$$

$$\psi_{qn} = exp\left(\tilde{\boldsymbol{\beta}}_{qn}'\tilde{\boldsymbol{z}}_{qn} + \tilde{\epsilon}_{qn}\right) \quad (19.2)$$

$$\psi_{qw} = exp\left(\tilde{\boldsymbol{\beta}}_{qw}'\tilde{\boldsymbol{z}}_{qw} + \tilde{\epsilon}_{qw}\right) \quad (19.3)$$

Applying the above stochastic specifications for baseline utility to the KKT conditions given in equations 14.1-14.6, and a ln() transformation, gives the following stochastic KKT conditions (for non-work activities)

$$ln\left(\frac{V_{qk}}{p_{qk}}\right)+\boldsymbol{\beta}'_{qk}\boldsymbol{z}_{qk}+\epsilon_{qk}=-ln\left(\frac{x^*_{q1}}{p_{q1}}\right)+\beta'_{q}z_{q1}+\epsilon_{q1} \quad \text{if } x^*_{qk}>0, k=2,3,\dots K \tag{20.1}$$

$$ln\left(\frac{V_{qk}}{p_{qk}}\right)+\boldsymbol{\beta}'_{qk}\boldsymbol{z}_{qk}+\epsilon_{qk}<-ln\left(\frac{x^*_{q1}}{p_{q1}}\right)+\boldsymbol{\beta}'_{q1}\boldsymbol{z}_{q1}+\epsilon_{q1} \quad \text{if } x^*_{qk}=0, k=2,3,\dots K \tag{20.2}$$

$$ln(\tilde{V}_{qn})+\widetilde{\boldsymbol{\beta}}'_{qn}\tilde{\boldsymbol{z}}_{qn}+\tilde{\epsilon}_{qn}=-ln(t^*_{q1})+\widetilde{\boldsymbol{\beta}}'_{qn}\tilde{\boldsymbol{z}}_{q1}+\tilde{\epsilon}_{q1} \quad \text{if } t^*_{qn}>0, n=2,3,\dots N \tag{20.3}$$

$$ln(\tilde{V}_{qn})+\widetilde{\boldsymbol{\beta}}'_{qn}\tilde{z}_{qn}+\tilde{\epsilon}_{qn}<-ln(t^*_{q1})+\widetilde{\boldsymbol{\beta}}'_{q1}\tilde{\boldsymbol{z}}_{q1}+\tilde{\epsilon}_{q1} \quad \text{if } t^*_{qn}=0, n=2,3,\dots N \tag{20.4}$$

and the additional stochastic KKT conditions for the work activity

$$ln(\tilde{V}_{qw})+\widetilde{\boldsymbol{\beta}}'_{qw}\tilde{\boldsymbol{z}}_{qw}+\tilde{\epsilon}_{qw}=ln\begin{pmatrix}(t^*_{q1})^{-1}\exp(\widetilde{\boldsymbol{\beta}}_{q}{}'\tilde{\boldsymbol{z}}_{q1}+\tilde{\epsilon}_{q1})\\ -\dfrac{\omega_w}{p_{q1}}(x^*_{q1})^{-1}\exp(\boldsymbol{\beta}'_{q1}\boldsymbol{z}_{q1}+\epsilon_{q1})\end{pmatrix} \quad \text{if } t^*_{qw}>0, w=1,2,\dots W \tag{20.5}$$

$$ln(\tilde{V}_{qw})+\widetilde{\boldsymbol{\beta}}'_{qw}\tilde{\boldsymbol{z}}_{qw}+\tilde{\epsilon}_{qw}<ln\begin{pmatrix}(t^*_{q1})^{-1}\exp(\widetilde{\boldsymbol{\beta}}'_{q1}\tilde{\boldsymbol{z}}_{q1}+\tilde{\epsilon}_{q1})\\ -\dfrac{\omega_q}{p_{q1}}(x^*_{q1})^{-1}\exp(\boldsymbol{\beta}'_{q1}\boldsymbol{z}_{q1}+\epsilon_{q1})\end{pmatrix} \quad \text{if } t^*_{qw}=0, w=1,2,\dots W \tag{20.6}$$

where

$$V_{qk}=\begin{cases}1 & \text{if } x^*_{qk}=0\\ \left(\dfrac{x^*_{qk}}{\gamma_k}+1\right)^{-1} & \text{otherwise}\end{cases} \tag{21.1}$$

$$V_{qn}=\begin{cases}1 & \text{if } t^*_{qn}=0\\ \left(\dfrac{t^*_{qn}}{\tilde{\gamma}_n}+1\right)^{-1} & \text{otherwise}\end{cases} \tag{21.2}$$

$$V_{qw}=\begin{cases}1 & \text{if } x^*_{qw}=0\\ \left(\dfrac{x^*_{qw}}{\tilde{\gamma}_k}+1\right)^{-1} & \text{otherwise}\end{cases} \tag{21.3}$$

To estimate the model, one of the baseline utility functions must be normalized (typically to 0) for consumption and time allocation. A simple estimation assumption is to normalize the baseline utilities for the outside goods. The likelihood function is further normalized by assuming $\sigma$ is equal to unity - a common practice. The double integral can be simplified by recognizing that each of the $k$ terms is independent of the error term $\tilde{\epsilon}_{q1}$ and each of the $n$ terms is independent of the error term $\epsilon_{q1}$. However, the work activity is dependent upon both error terms because it appears in both function constraints (i.e., has a multidimensional constraint). Following van Nostrand et al. (2013), it is assumed that the baseline utilities for both consumption and time allocation are the same for the outside goods. Given that both their baseline utilities contain only an error term, this assumption has almost no bearing on the likelihood function. Besides, only the work activity is affected by this assumption. These normalizations provide a means of maintaining a closed-form, which Astroza et al. (2017) do not maintain - resorting to the simulation of the conditional utility function.

The probability an individual consumes $M$ of the $k$ goods, $\widetilde{M}$ of the non-work activities, and the work activity is given by the following expression (assuming type I EV errors)[2]

$$P_q = \left[c_{qw}\prod_{k=2}^{K} c_{qk}\sum_{k=1}^{M}\frac{1}{c_{qk}}\prod_{n=2}^{\widetilde{M}} c_{qn}\sum_{n=1}^{\widetilde{M}}\frac{1}{c_{qn}}\right]\left[\frac{\widetilde{V}_{qw}}{a-b}\exp\left(\widetilde{\boldsymbol{\beta}}'_{qw}\widetilde{\boldsymbol{z}}_{qw}\right)\exp\left(-\frac{\widetilde{V}_{qw}}{a-b}\exp\left(\widetilde{\boldsymbol{\beta}}'_{qw}\widetilde{\boldsymbol{z}}_{qw}\right)\right)\right] \left[\frac{\prod_{M=1}^{M}\exp\left(W_{qk}\right)}{\sum_{k=1}^{K}\exp\left(W_{qk}\right)}(M-1)!\right]\left[\frac{\prod_{\widetilde{M}=1}^{\widetilde{M}}\exp\left(W_{qn}\right)}{\sum_{n=1}^{N}\exp\left(W_{qn}\right)}(\widetilde{M}-1)!\right] \quad (22)$$

where

$$c_{qk} = \frac{1}{x^*_{qk}+\widetilde{\gamma}_k} \quad (23.1)$$

$$c_{qn} = \frac{1}{t^*_{qn}+\widetilde{\gamma}_n} \quad (23.2)$$

$$c_{qw} = \frac{1}{t^*_{qw}+\widetilde{\gamma}_w} \quad (23.3)$$

$$W_{qk}|(\epsilon_{q1},\tilde{\epsilon}_{q1}) = -ln\left(\frac{x^*_{q1}}{p_{qk}}\right) - ln\left(\frac{V_{qk}}{p_{qk}}\right) - \boldsymbol{\beta}'_{qk}\boldsymbol{z}_{qk} + \epsilon_{q1} \quad (23.4)$$

$$W_{qn}|(\epsilon_{q1},\tilde{\epsilon}_{q1}) = -ln\left(t^*_{q1}\right) - ln\left(\widetilde{V}_{qn}\right) - \widetilde{\boldsymbol{\beta}}'_{qn}\widetilde{\boldsymbol{z}}_{qn} + \tilde{\epsilon}_{q1} \quad (23.5)$$

$$W_{qw}|(\epsilon^*_{q1}) = ln\left(\frac{\omega_q}{p_{q1}}(x^*_{q1})^{-1} - (t^*_{q1})^{-1}\right) - ln\left(\widetilde{V}_{qw}\right) - \widetilde{\boldsymbol{\beta}}'_{qw}\widetilde{\boldsymbol{z}}_{qw} + \epsilon^*_{q1} \quad (23.6)$$

$$a = \frac{\omega_q}{x^*_{q1}} \quad (23.7)$$

$$b = \frac{1}{t^*_{q1}} \quad (23.8)$$

The relationship between the *a* and *b* terms can be interpreted as stating that the ratio of wage over outside good consumption must be less than the dollar amount per hour spent on the outside good. In other words, if you have a good that you must consume, it must give you more value per unit time than working. For a valid solution, it must be true that $a - b > 0$.

---

[2] Additional details on the derivation of the likelihood function given in equation 22 are provided in an appendix.

Existing home production models make the strong assumption that all households contain only one person. This assumption arises from the challenge of allocating a household monetary budget among multiple members. A solution is proposed to this household budget allocation problem through a modification of the error structure of the likelihood function. The standard assumption in the MDCEV model is that errors are distributed as Type-I extreme value (as done in the above likelihood function derivation). However, it is possible to make an alternative assumption about the error structure and introduce correlations between alternatives and/or observations. For example, Pinjari and Bhat (2010) extend the MDCEV model to consider perfect substitution between a subset of alternatives via a re-specification of the error term to follow a nested GEV structure.

***Extension to a Household-Level Model***

To address the household budget allocation problem, the parallel constrained GEV structure of Gliebe and Koppelman (2005) is introduced to the MDCEV model. This paper incorporates this error structure into the home production function, and names this model as the parallel constrained MDCEV (PC-MDCEV) model. The optimization is then altered from one based on individuals to one based on households. Each member of the household has an independent temporal budget, which they are free to allocate to maximize their individual utility. The critical adjustment made in the PC-MDCEV model is that individuals are subject to a common parallel constraint among household members. The likelihood function for the home production function can then be extended from one based on persons to one based on households. Each member of the household has a temporal budget, while there is a single household monetary budget. The household model then has a time constraint for each member and a single household monetary budget constraint, which is represented as a parallel constraint acting on the utility maximization of each household member. The GEV structure of the parallel constrained model is defined by

$$G(Y_{11}, Y_{12}, \dots Y_{1k} \dots Y_{H1} \dots Y_{Hk}) = \sum_{b}^{B} \left[ \prod_{h}^{H} \left( \sum_{k}^{K} Y_{hk}^{\lambda_b} \right)^{\theta_h^q} \right]^{1/\lambda_b} \tag{24}$$

The original application of a parallel constrained GEV model was mode choice, where two individuals can choose a joint mode of travel. In such a case, the above GEV structure allows for individuals to choose a travel mode (e.g., drive or passenger) conditional upon a group choice of joint travel (e.g., carpool). This is captured through the specification of $B$ nests, representing the correlation of individual mode choices conditional upon a joint mode choice. The present application does not contain nests, and individuals are making the same choice of consumption. Therefore, the above equation can be simplified while maintaining the parallel constraint condition.

$$G(Y_{11}, Y_{12}, \dots Y_{1k} \dots Y_{H1} \dots Y_{Hk}) = \prod_{h}^{H} \left( \sum_{k}^{K} Y_{hk} \right)^{\theta_h^q} \tag{25}$$

It is assumed that each of the $h \in H$ members of the household make a decision about how much money the household should spend on each good and the contribution of each household member is given by $\theta_h^q$. The value of $\theta_h^q$ is a function of sociodemographic attributes describing individuals. The data generating function in equation 25 does not change the correlation structure between alternatives, and therefore, the MDCEV model structure is largely unchanged. The change comes in the discrete probability of consuming the set of $K$ goods, which is altered by the introduction of $\theta_h^q$. Assuming that $Y_{hk} = \exp(W_{hk})$, the household likelihood function for consumption of good $k$ is

$$P_{hk} = \frac{Y_{hk} G_{hk}(Y_{1k}, Y_{2k}, \dots Y_{Hk})}{\mu G(Y_{1k}, Y_{2k}, \dots Y_{Hk})} = \prod_{h}^{H} \frac{Y_{hk} G_{hk}(Y_{1k}, Y_{2k}, \dots Y_{Hk})}{\theta_h^q G(Y_{1k}, Y_{2k}, \dots Y_{Hk})} \tag{26}$$

Assume that $\mu = \sum_h^H \theta_h^q = 1$ to satisfy the GEV condition of homogeneity of degree one. While it should be noted that the function for each household member is not homogeneous of degree one, Ben-Akiva and Francois (1983) show that the homogeneity restriction can be relaxed to any degree of homogeneity. Again, the above condition does not affect the derivation of the MDCEV function. Rather, it simply re-weights $P_{qk}$ within each household, while maintaining a homogeneous of degree one MDCEV function for each $h$ household, as given in equation 26. Introducing the above changes to equation 22 gives the following parallel constrained individual utility function

$$P_q = \left[ c_{qw} \prod_{k=2}^{K} c_{qk} \sum_{k=1}^{M} \frac{1}{c_{qk}} \prod_{n=2}^{\widetilde{M}} c_{qn} \sum_{n=1}^{\widetilde{M}} \frac{1}{c_{qn}} \right] \left[ \frac{\widetilde{V}_{qw}}{a-b} \exp(\widetilde{\boldsymbol{\beta}}'_{qw} \widetilde{\mathbf{z}}_{qw}) \exp\left( -\frac{\widetilde{V}_{qw}}{a-b} \exp(\widetilde{\boldsymbol{\beta}}'_{qw} \widetilde{\mathbf{z}}_{qw}) \right) \right]$$
$$\prod_{k=1}^{M} \frac{exp(\theta_h^q W_{qk})}{\left( \sum_{k=1}^{K} \exp\left(\theta_h^q W_{qk}\right) \right)} (M-1)! \left[ \frac{\prod_{\widetilde{M}=1}^{\widetilde{M}} \exp\left(W_{qn}\right)}{\sum_{n=1}^{N} \exp\left(W_{qn}\right)} (\widetilde{M}-1)! \right] \tag{27}$$

To extend the model of individuals to households, we can define the following relationship

$$P_H = \prod_{h}^{H} P_{hk}^{\theta_h^q} P_{hn} P_{hw} \tag{28}$$

where $P_{hk} = P_{1k} = P_{2k} = P_{Hk}$ for each household member and $\theta_h^q$ denotes a weight for household member $h$ (or individual $q$). These weights represent the contribution of individual $h$'s utility in determining the household expenditure pattern. Given that households have different compositions, the allocation parameter $\theta_h^q$ can be parameterized as follows

$$\theta_h^q = \frac{\exp(\boldsymbol{\beta} \boldsymbol{Z}_h^q)}{\sum_h^H \exp(\boldsymbol{\beta} \boldsymbol{Z}_h^q)} \tag{29}$$

where $\boldsymbol{Z}_h^q$ is a set of individual characteristics for household member $h$ (or individual $q$) - age, employment status, or role in the household.

Finally, we can define a mapping matrix between individuals and households, which maps individual $q$ to its household. Transposing the matrix and summing across individuals gives the total utility for each household and provides the summation for the individual allocation equation. This allocation of income can be interpreted as providing an effective wage to household members who do not work. It is assumed that this is a household decision to maximize the utility of the household allocation between market and home production activities. The household members who can obtain the highest average effective wage will enter the market - that is, they will choose to work. This assumption is implicit in the time allocation observations but not explicitly represented in the model as a household-level negotiation. Note that this change in the likelihood function does not reflect a change in the optimization problem. Such a change would require an explicit consideration of intra-household negotiations that is beyond the scope of this paper. Zheng et al. (2021) provide a framework for considering household negotiations through an MDCEV time use task allocation model.

**Data Sources**

This section outlines the data, tested variables, and other details required to interpret the final model results presented in the following section. We start with a summary of the data construction procedure used in existing home production studies. A full description of our data construction and validation process can be found in Hawkins and Habib (2022).

Van Nostrand et al. (2013) develop a joint model for vacation decisions based on the American Time Use Survey and the U.S. Consumer Expenditure Survey. We use a similar set of data but a more robust data fusion procedure. First, the focus on vacations means that their model only requires travel time to the destination, time spent on vacation, and some measure of lodging and non-lodging costs. There is no need to distinguish between household members in their case as it is assumed that the entire family bears expenditures. Also, all household members have the same pattern of time use - all household members have the same travel time and spend the same amount of time on vacation.

In a series of studies, Jara-Diaz and colleagues estimate models of time use and expenditure for Karlsruhe, Germany; Santiago, Chile; Thurgau, Switzerland; the United States; the Netherlands; and Austria (Astroza et al., 2017; Hössinger et al., 2019; Jara-Díaz et al., 2008; Jokubauskaitė et al., 2019; Konduri et al., 2011). In the first three cases, the primary data source is a time use survey, and expenditures are coarsely linked through the sociodemographic characteristics of each household. The aggregate categorization of activities as travel, work, entertainment, and shopping means that a rough linkage is likely sufficient (Jara-Díaz et al., 2008). In another study, Konduri et al. (2011) fuse the American Time Use and Consumer Expenditure surveys along sociodemographic dimensions. However, in all their work, they reconcile the difference in the collection between time use (by the individual) and expenditure (by household) by only considering single-person households. In the final two studies (the Netherlands and Austria), surveys that combine time use and expenditure are available (Astroza et al., 2017; Hössinger et al., 2019). However, expenditures and time use are only collected as a diary, meaning that infrequent expenses are not collected from respondents. Also, it appears that only the time use information for the survey respondent was collected in both cases. Finally, the survey only includes working persons, as the model structure requires a wage rate. In our model framework, we overcome this limitation and therefore need information on all household members.

The primary data source for model estimation is a synthetic population for the Greater Toronto Area (GTA) constructed according to the process outlined in Figure 3 (and described in detail in a related paper). Data synthesis begins from a fusion along multiple dimensions of time use (TUS) and expenditure (SHS) surveys completed by Statistics Canada. Synthetic households are then constructed based on the household attributes provided by expenditure survey (SHS) respondents. We then address an unaddressed issue with such a data fusion exercise: fusion of survey records removes the spatial resolution critical to accurate analysis. Synthetic households are distributed to the study region (i.e., the GTA) using a combinatorial optimization population synthesis approach. This additional step helps to reintroduce spatial resolution by matching synthetic household attributes with marginal control totals for detailed spatial units (in our case aggregate dissemination areas). A random sample is drawn from this population for model estimation. After cleaning, the sample contains 9,625 individual records in 4,663 households. Land use variables are constructed from the 2016 Census of Population statistics (Statistics Canada, 2018) and DMTI enhanced points of interest (EPOI) data (DMTI, 2016). Land use statistics are calculated for these aggregate dissemination areas (ADAs).

Activities included in the model are work, home production, shopping, in-home food preparation and consumption, out-of-home food consumption, in-home entertainment, out-of-home entertainment, socializing with friends, and education. Home production time is defined as the aggregation of childcare; garbage, recycling, and unpacking goods; laundry, ironing, sewing, and shoe care; organizing, planning, and paying bills; packing/unpacking groceries, luggage, and boxes; pet care; and helping relatives, friends, neighbours, and acquaintances. Shopping is defined as time spent at the grocery store, other stores, or malls. In-home food time is defined as the aggregation of meal, lunch, and snack preparation; eating or drinking less time spent at a restaurant, bar, or club (defines out-of-food time); and preserving food. In-home entertainment time is defined as the aggregation of reading and television viewing. Out-of-home entertainment is defined as the aggregation of time at a sports center, field, arena, library, museum, or theatre. Socializing is defined as time spent socializing in-person, socializing using technology, organizational activities, and religious activities.

A similar set of expenditure categories are defined, with the removal of shopping because the purchase of non-food goods was considered irrelevant to the preference of home location choice. Dwelling expenditures (i.e., mortgage, rent, maintenance costs, and utilities) are included as an additional consumption category.

An extensive set of variables were tested for inclusion in the PC-MDCEV model, as summarized in Table 2

[Table 2]

Including interactions, the tested budget allocation variables are:

- Decadal age
- Youngest household member
- Oldest household member
- Sex
- Decadal age×Sex
- Marriage status
- Decadal age×Marriage status
- Marriage status×Dwelling type
- Main activity in last 7 days
- Workaholic
- Work balance
- Work from home
- Decadal age×Work balance

Several categorical variables were explored for inclusion in the model. The reference categories for these variables are summarized in Table 3.

[Table 3 here]

**Restrictions on Empirical Model relative to Theoretical Model**

Additional restrictions are placed on the model by practical and data considerations, which we summarize herein. The empirical application focuses on the estimation of the home production component of the model. Home location choice is considered in a parallel study (Hawkins et al., 2019).

Considering both location and home production faces a significant data challenge because the attributes of the non-chosen alternatives are unknown. The standard practice in the discrete choice estimation of home location choice models is to take a random sample of alternative locations, for which the attributes are known for the dwelling (e.g., number of bedrooms, number of bathrooms, etc.) and the location (e.g., distance to the workplace, distance to schools, etc.). In the proposed model, alternative home locations would make different time use and expenditure patterns available. Also, note that commuting time will affect the time available for work and forms another link between the monetary budget and time use. Unfortunately, there is no information on the individual time use and household consumption patterns for home locations that households have *not* chosen. Note, we are unaware of any existing model that proposes solution to the joint location and consumption model. Eliasson (2010) notes the limitation and includes only the results of an independently estimated mode choice model in his home location choice model. Martinez (2018) proposes an MDCEV model of consumption at alternative locations but offers no empirical application.

While the joint model is not estimated herein, we have a few suggestions for how it could be accomplished. One option would be to perform a stated adaptation experiment to ascertain how the consumption patterns of households are influenced by their location choice. Alternatively, an approach based on the work of Martinez (2018) and Eliasson (2010) is to independently estimate the home location and production components, then include an expectation term from the home production model in the location model. However, this expectation remains conditional upon home location. A market segmentation approach could be used to calculate aggregate market-specific expectation terms for each location. Finally, it would be plausible to use a propensity score matching approach to identify candidate households to represent time use and expenditure patterns at alternative locations. However, note that the preferences and budget restrictions of other households will influence both time allocation and expenditure patterns. As such, households from other locations may not represent the patterns of a given household if it chose that home location. There is also a self-selection effect in home location, such that alternative locations may not have suitable candidates (e.g., due to sociodemographic segregation within a region). With these notes, it is hoped that this initial empirical application can be extended by others to advance operational IUMs.

**Model Results**

The empirical model application is for the GTA, the largest metropolitan area in Canada with a 2021 population of 6.2 million people (Statistics Canada, 2022). The largest city in GTA, Toronto, has a 2021 population of 2.8 million people, or 7.6% of Canada's total population (City of Toronto, 2022). Mercer ranks Toronto as having the highest cost of living among Canadian cities and 89$^{th}$ worldwide (Mercer, 2023).

Table 4 presents the final specification of the model. A second model specification, with a slightly different specification (household size replaces marital status), is provided in Table 5 as an alternative specification. All baseline utility parameter values are significant at the 0.10 level except the ASC for social time. Versions of these models were also estimated with the inclusion of an alternative-specific constant (ASC) for work time. However, this specification encountered numerical convergence issues. Holding the satiation parameter fixed for work ($\gamma_w = 1.0$) and estimating the ASC resulted in similar convergence issues. The model was estimated in the GAUSS matrix programming language using code adapted from Bhat (2008).

Most satiation parameters are significant, which indicates that there are corner cases where individuals do not consume the good ($\gamma_1$ = 0 for outside goods because they are always consumed). Education has a large satiation parameter relative to other alternatives as there are a non-negligible number of individuals in the dataset who are currently enrolled in high school or university or for whom education constitutes a major activity. It has a high variation because most individuals over the age of 15 are not in school. Shopping has a significant non-zero satiation parameter, which indicates that the average individual enjoys this activity, although there may be heterogeneity within the population or between shopping categories (e.g., some people enjoy shopping for specific goods only). Out-of-home food production is viewed as leisure time relative to in-home food production – i.e., it has a larger satiation parameter indicating a stronger preference (or lower satiation).

***Effect of Household Demographics***

In line with labour survey statistics (Statistics Canada, 2007) the model suggests that female household members are less likely to spend time working than their male counterparts. Those who stated that they feel they have a good balance between work and leisure also tend to work less. Work parameters should be interpreted in light of the model, including all household members. For example, a negative sign for household size is due to larger households including non-working members and should not be interpreted to suggest that employed members of such households spend less time working. However, results do suggest that residents of the City of Toronto tend to spend more time working than residents of other regions. This result makes some sense since dwelling prices tend to be higher in Toronto, requiring additional work, additional workers, a higher wage, or some combination of these factors.

The results indicate that individuals in larger households are less likely to spend time on home production. This result represents an opportunity to apply the economics of the firm to an interpretation of household behaviour. Larger households, like larger firms, benefit from economies of scale. A limited number of household maintenance tasks can be spread among a larger number of individuals, with a diminishing household task burden with growing household size. For example, an additional household member may not increase the number of required loads of laundry - instead, each load may be larger. The positive sign for third income quartile time allocation to home production does not match the theoretical hypothesis of higher-income households spending less time on home production. Home production theory suggests that, as household income rises, households tend to spend less time on home production. Intermediate model specifications give a smaller and insignificant parameter value for the upper-income quartile relative to the third income quartile. This result suggests that there are unobserved variables not captured in the model. It may be the case that some lower-income households are working longer hours at a low wage rate and do not have as much time available for home production. It may also be the case that there is an interaction with household size - larger households have higher incomes due to the aggregation of individual incomes. This outcome supports the need for targeted econometric analysis of key variables to inform structural choice model estimation.

Age effects are significant in the time allocation component of the model. All age groups are less likely to spend time shopping than the youngest reference group. In contrast, the frequency of both in-home and out-of-home food consumption time tends to increase with age - younger individuals are in a rush to finish their meal and get on to other activities. In the case of out-of-home food, there may be an income effect as older individuals tend to have higher incomes. However, no significant income effect was found for out-of-home food in the consumption component of the model.

***Effect of Dwelling and Land Use Characteristics***

The interpretation of dwelling and land use parameters is less clear. In line with expectations, out-of-home food time is positively correlated with a higher density of office establishments. It is also correlated with a higher proportion of apartment dwellings, which does fit with the initial hypothesis that these households tend to eat out of the home rather than preparing their own meals. However, the model suggests that households living in ADAs with higher proportions of detached dwellings are less likely to allocate time to both in-home food and entertainment activities. There is no clear explanation for the cause of this effect. Perhaps, these households have longer commutes that leave them with little time to watch television or bake bread. Intermediate models also gave a negative parameter value for household size for these categories, which is correlated with living in a detached dwelling. These results suggest a need for continued analysis to determine the underlying drivers of such unexpected parameter signs.

Regarding consumption, households in ADAs with higher proportions of detached dwellings (i.e., suburbs) are more likely to spend money on substitution of home production tasks. Again, the hypothesis was that households in high-density areas tend to substitute away from home production time. There are several explanations for a higher consumption of home production services by households living in detached dwellings. They may pay more for these services (housekeepers, cleaners, and house-sitters) because of the larger size of dwellings. Larger detached dwellings, compared with apartments and townhouses, are more expensive, and therefore, their owners may have a higher income, which allows for greater consumption of these services. It may be the case that some of these households are still in urban areas, but that their ADAs are characterized by higher proportions of detached dwellings. This hypothesis is borne out by the positive parameter signs for in-home food, out-of-home food, in-home entertainment, and out-of-home entertainment. Households in ADAs with higher median incomes are more likely to spend on education. Finally, households in areas with high office establishment densities tend to have lower mortgages and rents. This effect is likely a function of the area of dwellings being smaller in these ADAs.

***Household Budget Allocation***

The final component of the model to be discussed is the allocation of the household monetary budget. As discussed above, several variable combinations were explored for this component of the model. The variables that appear in the final model consistently arose as significant in most intermediate models. Married females tend to have less influence on the allocation of the household budget than their male counterparts. However, females in the age group 35 to 44 tend to have a stronger influence on the budget than other age-sex combinations. In contrast, females in the age group 65 to 74 tend to have a lower-than-average influence on the budget, which is consistent with observations of growing equality across generations (Amato et al., 2003; Cineli, 2020). Finally, household members who consider themselves workaholics and are in the age group 35 to 44 tend to spend less of the household budget than other groups. These individuals may spend more time working and therefore may not have the necessary time to consume leisure goods and services at the same rate as other members of their household.

*Value of Time*

The value of leisure (VL), or time as a resource, and value of work (VW) can be calculated from the ratio of the Lagrangian multipliers and wage rate as follows:

$$VL = \frac{\mu_q}{\lambda_q} = \frac{\tilde{u}_w'\left(t_{qw}^*\right)}{u_1{}'\left(x_{q1}^*\right)} + \omega_q \tag{30.1}$$

$$VW = \frac{\mu_q}{\lambda_q} = \frac{\tilde{u}_w'\left(t_{qw}^*\right)}{u_1{}'\left(x_{q1}^*\right)} \tag{30.2}$$

The average values obtained from the model are \$57.59 per hour and \$16.82 per hour, respectively. While there are differences in the model specification, a comparison can be made with the results of Astroza et al. (2017) to confirm the magnitude of these values. They find similar values of €36.20 (\$54.23) per hour and €18.90 (\$28.32) per hour, respectively.

## Conclusions and Future Research

This paper provides a theoretical basis for transportation and land use system integration founded home production that can capture both the temporal and monetary constraints placed on households. The model represents multi-person households through the introduction of a parallel constraint on the household monetary budget. The model is estimated for a nine time and eight consumption category choice system. Various individual, household, dwelling, and neighborhood attributes are considered for inclusion in the model. The final model provides several novel results, including identifying an economy of scale in time devoted to home production. It was found that the mix of dwelling types (detached, apartment, etc.) has a significant influence on both time use and consumption. However, the interpretation of several of these effects requires additional study. A portion of these non-intuitive results may also be attributable to the synthetic dataset. The PC-MDCEV model structure can provide a detailed representation of intra-household budget allocation. Like any good scientific endeavour, the model and analysis outlined in this paper provide several advancements and new insights but raise just as many questions.

Additional applications of the PC-MDCEV models are necessary to validate the model parameters, explore the inclusion of additional variables, and test the sensitivity of the model results to the specification of the input data. While our data synthesis process is more robust than that used in extant home production function estimations, questions remain regarding the impact of synthesis on model parameter signs and magnitudes in the resulting model. In other research, we present reduced-form analysis as a means of isolating components of the structural model (Hawkins and Habib, 2021). Additional causal inference of time use and consumption patterns would better inform the interpretation of the PC-MDCEV model results.

Bhat has proposed changes to the basic MDCEV structure and accommodation of multiple constraints that could be incorporated into the PC-MDCEV model. In the first instance, he estimates separate baseline utility functions for the discrete and continuous components (2018). In a second paper, Mondal and Bhat (2020) reformulate the objective function as a minimization (rather than a maximization), which provides a closed-form MDCEV structure that can more easily accommodate multiple constraints.

The PC-MDCEV model presented in the current paper relies on an assumption of independence between the time budgets of household members. This limitation should be addressed in subsequent work to recognize household trade-offs and cooperation in completing household activities. There is a rich literature in the home economics literature on this topic (Browning et al., 1994; Browning and Chiappori, 1998; Chiappori, 1997; Chiappori et al., 2014; Chiappori and Mazzocco, 2017) and Jara-Diaz and Candia (2020) provide a transportation application. As referenced above, Zheng et al. (2021) provide a first effort in this direction by modeling task allocation among elderly couples in Beijing, China. They also start from an MDCEV foundation, making integration with our model relatively straightforward. However, their dataset is a household travel diary with household task allocation details not available in our dataset. Integration with our contribution would require additional data collection.

Alternative specifications could address the homothetic preferences assumption of the LES utility specification. The homotheticity assumption is an important limitation of the current MDCEV specification, as it ignores the effect of income on consumption patterns. Bhat et al. (2015) explore using a utility specification developed by Lavin and Hanemann (2008) to address the problem, but there is much work remaining in this area. Other standard demand specifications could be considered, such as the translog demand system of Christensen et al. (1975) or the almost ideal demand system (AIDS) of Deaton and Muellbauer (1980). However, these changes would require the derivation of an updated MDCEV model structure and would likely introduce additional estimation challenges.

The multiple discrete-continuous extreme value (MDCEV) model represents an optimization of the discrete choice among alternatives and the continuous choice of consumption. The inclusion of this optimization introduces a significant challenge to simulation relative to the discrete choice models used in most transport demand models. The model simulation requires the application of this optimization for every individual in the full population, a computationally burdensome task. However, in the current context, the objective is to capture the patterns of expected time use and consumption rather than to simulate each choice. We are interested in the results only as a means of characterizing accessibility and expenditure patterns. We propose using Bayesian inference to overcome the challenges of applying the model in simulation. The full posterior distribution can be computed using a sample of households. Draws can then be taken from the resulting posterior to characterize expectations for the full population of households. Bhat (2008) proposes the prediction of expenditures through the averaging of simulation draws (e.g., 5000 random draws) from the Gumbel distribution. Constrained optimization could then be used to obtain the predicted time use and expenditure values. In a subsequent study, Pinjari and Bhat (2021) outline a computationally efficient procedure for performing prediction that avoids the need for constrained optimization. In an empirical application, they find it takes almost four hours to perform a single set of draws for 4,382 individuals with the original method, suggesting it would take ~15 days to perform 100 draws for this sample. Their algorithm dramatically improves the time to simulate results. However, as an alternative, we suggest a Bayesian approach, as employed in the marketing literature (Allenby et al., 2017). This approach removes the need to perform multiple draws and average the results because the estimation step implicitly includes constructing a posterior distribution. The modeler can use the demographic and land use variables of out-of-sample households to perform simulation draws of their consumption patterns from the posterior distribution.

**Funding details**

The study was funded by an NSERC CGS-D Scholarship and a CRDCN Emerging Scholar Grant by the first author and an NSERC Discovery Grant by the second author.

## Disclosure statement

On behalf of all authors, the corresponding author states that there is no conflict of interest.

**Appendix: Additional Details on the Derivation of the MDCEV Home Production Model**

In the following likelihood function, $f(.)$ is the Type-1 extreme value PDF and $F(\ )$ is its corresponding CDF. The probability that a member $h$ of household $H$ consumes $M$ of the $k$ goods, $\widetilde{M}$ of the non-work activities, and the work activity is given by

$$
\begin{aligned}
& P_q\left(x_{q1}^*, x_{q2}^*, \dots x_{qM}^*, 0,0 \dots, 0, t_{q1}^*, t_{q2}^*, \dots t_{q\widetilde{M}}^*, 0,0, \dots 0, t_{qw}^*\right) \\
& = \frac{1}{\sigma\sigma^{M-1}\sigma^{\widetilde{M}-1}}\left[c_{qw}\prod_{k=2}^{K} c_{qk}\sum_{k=1}^{M}\frac{1}{c_{qk}}\prod_{n}^{\widetilde{M}} c_{qn}\sum_{k=1}^{\widetilde{M}}\frac{1}{c_{qn}}\right] \\
& \int_{-\infty}^{+\infty}\int_{-\infty}^{+\infty}\left[\prod_{k=2}^{M} f\left(\frac{W_{qk}|(\epsilon_{q1},\tilde{\epsilon}_{q1})}{\sigma}\right)\prod_{k=2}^{M} F\left(\frac{W_{qk}|(\epsilon_{q1},\tilde{\epsilon}_{q1})}{\sigma}\right)\right] \\
& \left[f\left(\frac{W_{qn}|(\epsilon_{q1},\tilde{\epsilon}_{q1})}{\sigma}\right)\prod_{n=2}^{\widetilde{M}} F\left(\frac{W_{qn}|(\epsilon_{q1},\tilde{\epsilon}_{q1})}{\sigma}\right)\right] \\
& \left[f\left(\frac{W_{qw}|(\epsilon_{q1},\tilde{\epsilon}_{q1})}{\sigma}\right)F\left(\frac{W_{qw}|(\epsilon_{q1},\tilde{\epsilon}_{q1})}{\sigma}\right)\right] f(\epsilon_{q1})f(\tilde{\epsilon}_{q1})d\epsilon_{q1}d\tilde{\epsilon}_{q1}
\end{aligned}
\qquad (A1)
$$

where

$$c_{qk} = \frac{1}{x_{qk}^* - x_{qk}^0 + \gamma_k} \qquad (A2.1)$$

$$c_{qn} = \frac{1}{t_{qn}^* - t_{qn}^0 + \tilde{\gamma}_n} \qquad (A2.2)$$

$$c_{qw} = \frac{1}{t_{qw}^* - t_{qw}^0 + \tilde{\gamma}_w} \qquad (A2.3)$$

$$W_{qk}|(\epsilon_{q1},\tilde{\epsilon}_{q1}) = -ln\left(\frac{x_{q1}^* - x_{qk}^0}{p_{qk}}\right) - ln\left(\frac{V_{qk}}{p_{qk}}\right) - \boldsymbol{\beta}'_{qk}\mathbf{z}_{qk} + \epsilon_{q1} \qquad (A2.4)$$

$$W_{qn}|(\epsilon_{q1},\tilde{\epsilon}_{q1}) = -ln\left(t_{q1}^* - t_{q1}^0\right) - ln\left(\tilde{V}_{qn}\right) - \tilde{\boldsymbol{\beta}}'_{qn}\mathbf{z}_{qn} + \tilde{\epsilon}_{q1} \qquad (A2.5)$$

$$W_{qw}|(\epsilon_{q1},\tilde{\epsilon}_{q1}) = ln\left((t_{q1}^* - t_{q1}^0)^{-1}\exp(\tilde{\epsilon}_{q1}) - \frac{\omega_w}{p_{q1}}(x_{q1}^* - x_{q1}^0)^{-1}\exp(\epsilon_{q1})\right) - ln\left(\tilde{V}_{qw}\right) - \tilde{\boldsymbol{\beta}}'_{qw}\tilde{\mathbf{z}}_{qw} + \epsilon_{q1}^* \qquad (A2.6)$$

We confirmed the accuracy of our formulation by checking the determinant value obtained with synthetic data against a built-in determinant function in GAUSS (Aptech, 2018). With the assumption $\epsilon_{q1}^* = \epsilon_{q1} = \tilde{\epsilon}_{q1}$, the probability can be simplified to

$$P_q = \left[c_{qw}\prod_{k=2}^{K} c_{qk}\sum_{k=1}^{M}\frac{1}{c_{qk}}\prod_{n=2}^{\tilde{M}} c_{qn}\sum_{n=1}^{\tilde{M}}\frac{1}{c_{qn}}\right]\left[\frac{\prod_{M=1}^{M}\exp\left(W_{qk}\right)}{\sum_{k=1}^{K}\exp\left(W_{qk}\right)}(M-1)!\right]$$
$$\left[\frac{\prod_{\tilde{M}=1}^{\tilde{M}}\exp\left(W_{qn}\right)}{\sum_{n=1}^{N}\exp\left(W_{qn}\right)}(\tilde{M}-1)!\right]\int_{-\infty}^{+\infty}\exp\left(-W_{qw}\right)\exp\left(-\exp\left(-W_{qw}\right)\right)\exp\left(-\epsilon_{q1}^{*}\right) \qquad (A3)$$

where $W_{qw}$ is now given by

$$W_{qw}|(\epsilon_{q1}^{*}) = ln\left(\frac{\omega_q}{p_{q1}}(x_{q1}^{*}-x_{q1}^{0})^{-1} - (t_{q1}^{*}-t_{q1}^{0})^{-1}\right) - ln\left(\tilde{V}_{qw}\right) - \tilde{\boldsymbol{\beta}}'_{qw}\tilde{\mathbf{z}}_{qw} + \epsilon_{q1}^{*} \qquad (A4)$$

The $\sigma$s are suppressed for ease of interpretation and because each is normalized to one. The sub-component of the likelihood function for work time allocation can be written as

$$P_{qw} = c_{qw}\int_{-\infty}^{+\infty}\exp\left(-\ln(a-b) + \ln\left(\tilde{V}_{qw}\right) + \tilde{\boldsymbol{\beta}}'_{qw}\tilde{\mathbf{z}}_{qw}\right)\exp\left(-\epsilon_{q1}^{*}\right)$$
$$\exp\left(-\exp\left(-\ln(a-b) + ln\left(\tilde{V}_{qw}\right) + \tilde{\boldsymbol{\beta}}'_{qw}\tilde{\mathbf{z}}_{qw}\right)\right)\exp\left(-\exp\left(-\epsilon_{q1}^{*}\right)\right)\exp\left(-\epsilon_{q1}^{*}\right)d\epsilon_{q1}^{*} \qquad (A5)$$

where

$$a = \frac{\omega_q}{x_{q1}^{*} - x_{q1}^{0}} \qquad (A6.1)$$
$$b = \frac{1}{t_{q1}^{*} - t_{q1}^{0}} \qquad (A6.2)$$

The relationship between the *a* and *b* terms can be interpreted as stating that the ratio of wage over outside good consumption must be less than the dollar amount per hour spent on the outside good. In other words, if you have a good that you must consume, it must give you more value per unit time than working.

The above expression can be simplified to

$$P_{qw} = c_{qw}\frac{\tilde{V}_{qw}}{a-b}\exp\left(\tilde{\boldsymbol{\beta}}'_{qw}\tilde{\mathbf{z}}_{qw}\right)\exp\left(-\frac{\tilde{V}_{qw}}{a-b}\exp\left(\tilde{\boldsymbol{\beta}}'_{qw}\tilde{\mathbf{z}}_{qw}\right)\right)$$
$$\int_{-\infty}^{+\infty}\exp\left(-\epsilon_{q1}^{*}\right)\exp\left(-\exp\left(-\epsilon_{q1}^{*}\right)\right)\exp\left(-\epsilon_{q1}^{*}\right)d\epsilon_{q1}^{*} \qquad (A7)$$

With the substitutions $r = \exp(-\epsilon_{q1}^*)$ and $dr = -\exp(-\epsilon_{q1}^*)d\epsilon_{q1}^*$, the integration can be written as

$$P_{qw} = c_{qw}\frac{\tilde{V}_{qw}}{a-b}\exp(\widetilde{\boldsymbol{\beta}}'_{qw}\tilde{\mathbf{z}}_{qw})\exp\left(-\frac{\tilde{V}_{qw}}{a-b}\exp(\widetilde{\boldsymbol{\beta}}'_{qw}\tilde{\mathbf{z}}_{qw})\right)\int_{\infty}^{0} -r\exp(-r)dr \qquad (A8)$$

Integration by parts can then be employed to solve the integral where $u = r, du = 1, dv = -\exp(-r)$, and $v = \exp(-r)$

$$P_{qw} = c_{qw}\frac{\tilde{V}_{qw}}{a-b}\exp(\widetilde{\boldsymbol{\beta}}'_{qw}\tilde{\mathbf{z}}_{qw})\exp\left(-\frac{\tilde{V}_{qw}}{a-b}\exp(\widetilde{\boldsymbol{\beta}}'_{qw}\tilde{\mathbf{z}}_{qw})\right)\left((r+1)\exp^{-r}|_{\infty}^{\circ}\right) \qquad (A9)$$

This can be simplified to

$$P_{qw} = c_{qw}\frac{\tilde{V}_{qw}}{a-b}\exp(\widetilde{\boldsymbol{\beta}}'_{qw}\tilde{\mathbf{z}}_{qw})\exp\left(-\frac{\tilde{V}_{qw}}{a-b}\exp(\widetilde{\boldsymbol{\beta}}'_{qw}\tilde{\mathbf{z}}_{qw})\right) \qquad (A10)$$

**Table 1** Summary of Notation Used in PC-MDCEV Home Production Model

| | |
|---|---|
| $U_q(\ \ )$ | Quasi-concave, increasing, and continuously differentiable direct utility function with respect to the consumption quantities for individual $q$ |
| $E_q$ | Monetary endowment beyond hourly wages (e.g., retirement savings, gifts, etc.) for individual $q$ |
| $T_q$ | Total time available (i.e., 24-hours per day) for individual $q$ |
| $u_k(x_{qk})$ | Direct utility component for good $k$ consumed by individual $q$ |
| $\tilde{u}_n(t_{nq})$ | Direct utility component for non-work activity $n$ time use by individual $q$ |
| $\tilde{u}_w(t_{wq})$ | Direct utility component for work activity $w$ time use by individual $q$ |
| $p_{qk}$ | Unit price for good $k$ consumed by individual $q$ (assumed unity, 1.0) |
| $x_{qk}$ | Good $k$ quantity consumed by individual $q$ |
| $t_{qn}$ | Non-work activity $n$ time use by individual $q$ |
| $t_{qw}$ | Work activity $w$ time use by individual $q$ |
| $\psi_{qk}, \tilde{\psi}_{qn}, \tilde{\psi}_{qw}$ | Baseline marginal utility for good $k$ consumption by individual $q$, where $k$ for goods consumption is replaced by $n$ for non-work time use and $w$ for work time use |
| $\gamma_k, \tilde{\gamma}_n, \tilde{\gamma}_w$ | Translation (satiation) parameter for good $k$, where $k$ for goods consumption is replaced by $n$ for non-work time use and $w$ for work time use |
| $\lambda_q$ | Lagrangian multiplier for monetary budget constraint for individual $q$ |
| $\mu_q$ | Lagrangian multiplier for temporal budget constraint for individual $q$ |
| $\boldsymbol{\beta}_{qk}, \widetilde{\boldsymbol{\beta}}_{qn}, \widetilde{\boldsymbol{\beta}}_{qw}$ | Vector of parameters associated with variables included in baseline marginal utility function for good $k$ consumption by individual $q$, where $k$ for goods consumption is replaced by $n$ for non-work time use and $w$ for work time use |
| $\boldsymbol{z}_{qk}, \tilde{\mathbf{z}}_{qn}, \tilde{\mathbf{z}}_{qw}$ | Vector of variables included in baseline marginal utility function for good $k$ consumption by individual $q$, where $k$ for goods consumption is replaced by $n$ for non-work time use and $w$ for work time use |
| $\boldsymbol{Z}_h^q$ | Individual characteristics for household member $h$ (or individual $q$) used in household utility weighting function |
| | |

**Table 2** Summary of Considered Model Variables

| Variable name | Variable Description |
|---|---|
| **Individual Variables** | |
| Decadal age | Individual age measured in 10-year increments |
| Sex | Male or female |
| Marital status | Never married, married or living common law, divorced/widowed/separated |
| Occupation type | 10 categories |
| Time in job | Years |
| Workaholic | Do you consider yourself a workaholic? |
| Work balance | Satisfaction on 5-point scale |
| Work from home | Excluding overtime, do you usually work any of your scheduled hours at home? |
| Main activity in last 7 days | 10 categories |
| **Household Variables** | |
| Children in household | Yes/no |
| Household type | 7 categories |
| Income | |
| **Dwelling Variables** | |
| Dwelling type | 6 categories |
| Year built | |
| **Land Use Variables** | |
| Region | Durham, Halton, Peel, Toronto, or York |
| Population density | Persons per ha |
| Office density | Establishments per ha |
| Retail density | Establishments per 1000 persons |
| Entropy of establishments in ADA | As defined in Piovani (2018) |
| Median income in ADA | Measured in $10,000 |
| Proportion of commutes in ADA by car | |
| Proportion of commutes in ADA by active modes | |
| Proportion of owned dwellings in ADA | |
| Proportion of detached dwellings in ADA | |
| Proportion of apartment dwellings in ADA | |
| Proportion of dwellings in ADA built before 1960 | |
| Proportion of dwellings in ADA built between 1960 and 2000 | |

**Table 3** Categorical Variable Reference Values

| Variable | Reference Category |
|---|---|
| Age of individual | ≥75 |
| Individual main activity in last 7 days | Working |
| Dwelling type | Single-family detached |
| Household type | Single unmarried |
| Household income quartile | Lower quartile |
| Home region | City of Toronto |
| Dwelling year built | After 2000 |
| Marriage status | Single unmarried |

**Table 4** PC-MDCEV Model Results with Marital status in Baseline Utility of Expenditure

| | **Variable** | **(1)** | **(2)** | **(3)** | **(4)** | **(5)** | **(6)** | **(7)** | **(8)** | **(9)** | **(10)** |
|---|---|---|---|---|---|---|---|---|---|---|---|
| Time Use | | **Baseline Marginal Utility (β)** | | | | | | | | | |
| | ASC | - | 2.12 | -4.26 | 1.40 | -4.23 | -0.44 | -3.90 | 0.11 | - | -8.36 |
| | Female | -0.76 | - | - | - | - | - | - | - | - | - |
| | Age <25 | - | | | 0.36 | 0.43 | - | - | - | - | |
| | Age 25 to 34 | - | -0.59 | -0.29 | 0.44 | 0.50 | - | - | - | - | -0.65 |
| | Age 35 to 44 | - | -0.25 | -0.20 | 0.45 | 0.47 | - | -0.10 | - | - | -2.19 |
| | Age 45 to 54 | - | 0.29 | -0.37 | 0.50 | 0.53 | 0.21 | | -0.12 | - | -2.76 |
| | Age 55 to 64 | - | - | -0.42 | 0.66 | 0.56 | 0.46 | -0.49 | -0.40 | - | - |
| | Age 65 to 74 | - | - | -0.35 | 0.62 | 0.53 | 0.47 | -0.19 | -0.51 | - | - |
| | Work balance (balanced) | -1.11 | - | - | - | - | - | - | - | - | - |
| | HH size | -1.85 | -0.39 | - | - | - | - | - | - | - | - |
| | HH type (family) | -2.34 | - | - | - | - | - | - | - | - | - |
| | HH income (3rd quartile) | - | 0.19 | - | 0.42 | - | - | - | - | - | - |
| | Durham region | -0.58 | - | - | 0.22 | 0.51 | - | - | - | - | - |
| | York region | -1.21 | - | - | - | - | - | -0.79 | - | - | - |
| | Peel region | - | - | -0.53 | - | - | - | -0.82 | - | - | - |
| | Halton region | -1.43 | - | - | 0.50 | 0.36 | 0.22 | - | - | - | - |
| | Proportion dwellings in ADA (detached) | - | - | - | -0.14 | - | -0.14 | - | - | - | - |
| | Proportion dwellings in ADA (apartment) | - | - | 0.18 | - | 0.15 | - | - | - | - | - |
| | Proportion of commuters in ADA (walk+bike) | - | 0.41 | - | - | - | - | - | - | - | - |
| | Office establishment density in ADA (per km$^2$) | - | - | - | - | 0.02 | - | - | - | - | - |
| | | **Satiation parameter (γ)** | | | | | | | | | |
| | | 0.02 | 0.06 | 1.40 | 0.02 | 0.92 | 0.11 | 0.54 | 0.11 | - | 124.80 |
| | | **Baseline Marginal Utility (β)** | | | | | | | | | |
| Expenditure | ASC | - | -0.94 | - | 3.63 | 0.65 | 3.75 | 1.67 | - | 6.64 | -1.25 |
| | HH size | - | -0.28 | - | -0.33 | -0.27 | -0.26 | -0.20 | - | -0.32 | -0.30 |
| | Townhouse/semi-detached dwelling | - | - | - | 0.61 | 0.17 | 0.43 | 0.29 | - | 0.58 | 0.23 |
| | Apartment dwelling | - | - | - | 0.94 | - | 0.58 | 0.30 | - | 1.02 | - |
| | Proportion dwellings in ADA (detached) | - | 1.66 | - | 0.48 | 0.53 | 0.34 | 0.68 | - | - | 0.25 |
| | Proportion dwellings in ADA (apartment) | - | - | - | - | - | - | - | - | - | -0.39 |
| | Median income in ADA | - | - | - | - | - | - | - | - | - | 0.03 |
| | Proportion of commuters in ADA (walk+bike) | - | - | - | - | - | - | - | - | - | -1.08 |
| | Office establishment density in ADA (per km$^2$) | - | - | - | - | - | - | - | - | -0.03 | - |
| | | **Satiation parameter (γ)** | | | | | | | | | |
| | | - | 0.01 | - | 0.01 | 0.02 | 0.00 | 0.02 | - | 0.00 | 0.04 |
| | | **Individual Weights** | | | | | | | | | |
| Budget | Married and female | -0.18 | | | | | | | | | |
| | Age 35 to 44 and female | 0.18 | | | | | | | | | |
| | Age 65 to 74 and female | -0.27 | | | | | | | | | |

|  |  |  |
|---|---|---|
|  | Age 35 to 44 and workaholic | -0.21 |
|  | AIC | 281157.8 |

**Note:** (1) = Work, (2) = Home Production, (3) = Shopping, (4) = In-Home Food Production, (5) = Out-Of-Home Food Production, (6) = In-Home Entertainment Production, (7) = Out-Of-Home Entertainment Production, (8) = Social, (9) = Dwelling, (10) = Education

**Table 5** PC-MDCEV Model Results with Household Size in Expenditure

| | **Variable** | **(1)** | **(2)** | **(3)** | **(4)** | **(5)** | **(6)** | **(7)** | **(8)** | **(9)** | **(10)** |
|---|---|---|---|---|---|---|---|---|---|---|---|
| Time Allocation | | **Baseline Marginal Utility (β)** | | | | | | | | | |
| | ASC | - | 2.12 | -4.26 | 1.40 | -4.23 | -0.44 | -3.90 | 0.11 | - | -8.36 |
| | Female | -0.76 | - | - | - | - | - | - | - | - | - |
| | Age <25 | - | | | 0.36 | 0.43 | - | - | - | - | |
| | Age 25 to 34 | - | -0.59 | -0.29 | 0.44 | 0.50 | - | - | - | - | -0.65 |
| | Age 35 to 44 | - | -0.25 | -0.20 | 0.45 | 0.47 | - | -0.10 | - | - | -2.19 |
| | Age 45 to 54 | - | 0.29 | -0.37 | 0.50 | 0.53 | 0.21 | | -0.12 | - | -2.76 |
| | Age 55 to 64 | - | - | -0.42 | 0.66 | 0.56 | 0.46 | -0.49 | -0.40 | - | - |
| | Age 65 to 74 | - | - | -0.35 | 0.62 | 0.53 | 0.47 | -0.19 | -0.51 | - | - |
| | Work balance (balanced) | -1.11 | - | - | - | - | - | - | - | - | - |
| | HH size | -2.35 | -0.39 | - | - | - | - | - | - | - | - |
| | HH type (family) | -2.34 | - | - | - | - | - | - | - | - | - |
| | HH income (3rd quartile) | - | 0.19 | - | 0.42 | - | - | - | - | - | - |
| | Durham region | -0.58 | - | - | 0.22 | 0.51 | - | - | - | - | - |
| | York region | -1.21 | - | - | - | - | - | -0.79 | - | - | - |
| | Peel region | - | - | -0.53 | - | - | - | -0.82 | - | - | - |
| | Halton region | -1.43 | - | - | 0.50 | 0.36 | 0.22 | - | - | - | - |
| | Proportion dwellings in ADA (detached) | - | - | - | -0.14 | - | -0.14 | - | - | - | - |
| | Proportion dwellings in ADA (apartment) | - | - | 0.18 | - | 0.15 | - | - | - | - | - |
| | Proportion of commuters in ADA (walk+bike) | - | 0.41 | - | - | - | - | - | - | - | - |
| | Office establishment density in ADA (per $km^2$) | - | - | - | - | 0.02 | - | - | - | - | - |
| | | **Satiation parameter (γ)** | | | | | | | | | |
| | | 0.02 | 0.06 | 1.40 | 0.02 | 0.92 | 0.11 | 0.54 | 0.11 | - | 124.71 |

| | | Baseline Marginal Utility (β) | | | | | | | | | |
|---|---|---|---|---|---|---|---|---|---|---|---|
| Consumption | ASC | - | -1.92 | - | 2.49 | -0.29 | 2.84 | 0.94 | - | 5.55 | -2.34 |
| | Married | - | 0.84 | - | 1.04 | 0.66 | 0.76 | 0.60 | - | 0.95 | 0.85 |
| | Divorced/separated/widowed | - | 0.71 | - | 1.08 | 0.73 | 0.84 | 0.61 | - | 1.03 | 1.05 |
| | Townhouse/semi-detached dwelling | - | - | - | 0.55 | 0.18 | 0.40 | 0.29 | - | 0.54 | 0.23 |
| | Apartment dwelling | - | - | - | 0.76 | - | 0.49 | 0.26 | - | 0.89 | - |
| | Proportion dwellings in ADA (detached) | - | 1.69 | - | 0.49 | 0.54 | 0.36 | 0.71 | - | - | 0.32 |
| | Proportion dwellings in ADA (apartment) | - | - | - | - | - | - | - | - | - | -0.39 |
| | Median income in ADA | - | - | - | - | - | - | - | - | - | 0.03 |
| | Proportion of commuters in ADA (walk+bike) | - | - | - | - | - | - | - | - | - | -1.07 |
| | Office establishment density in ADA (per km$^2$) | - | - | - | - | - | - | - | - | -0.04 | - |
| | | Satiation parameter (γ) | | | | | | | | | |
| | | - | 0.01 | - | 0.02 | 0.00 | 0.02 | 0.04 | - | 0.00 | 0.01 |
| | | Individual Weights | | | | | | | | | |
| Budget Allocation | Married and female | -0.23 | | | | | | | | | |
| | Age 35 to 44 and female | 0.18 | | | | | | | | | |

| | | |
|---|---|---|
| | Age 65 to 74 and female | -0.24 |
| | Age 35 to 44 and workaholic | -0.20 |
| | AIC | 281155.9 |

***Note:*** *(1) = Work, (2) = Home Production, (3) = Shopping, (4) = In-Home Food Production, (5) = Out-Of-Home Food Production, (6) = In-Home Entertainment Production, (7) = Out-Of-Home Entertainment Production, (8) = Social, (9) = Dwelling, (10) = Education*

## List of Figures

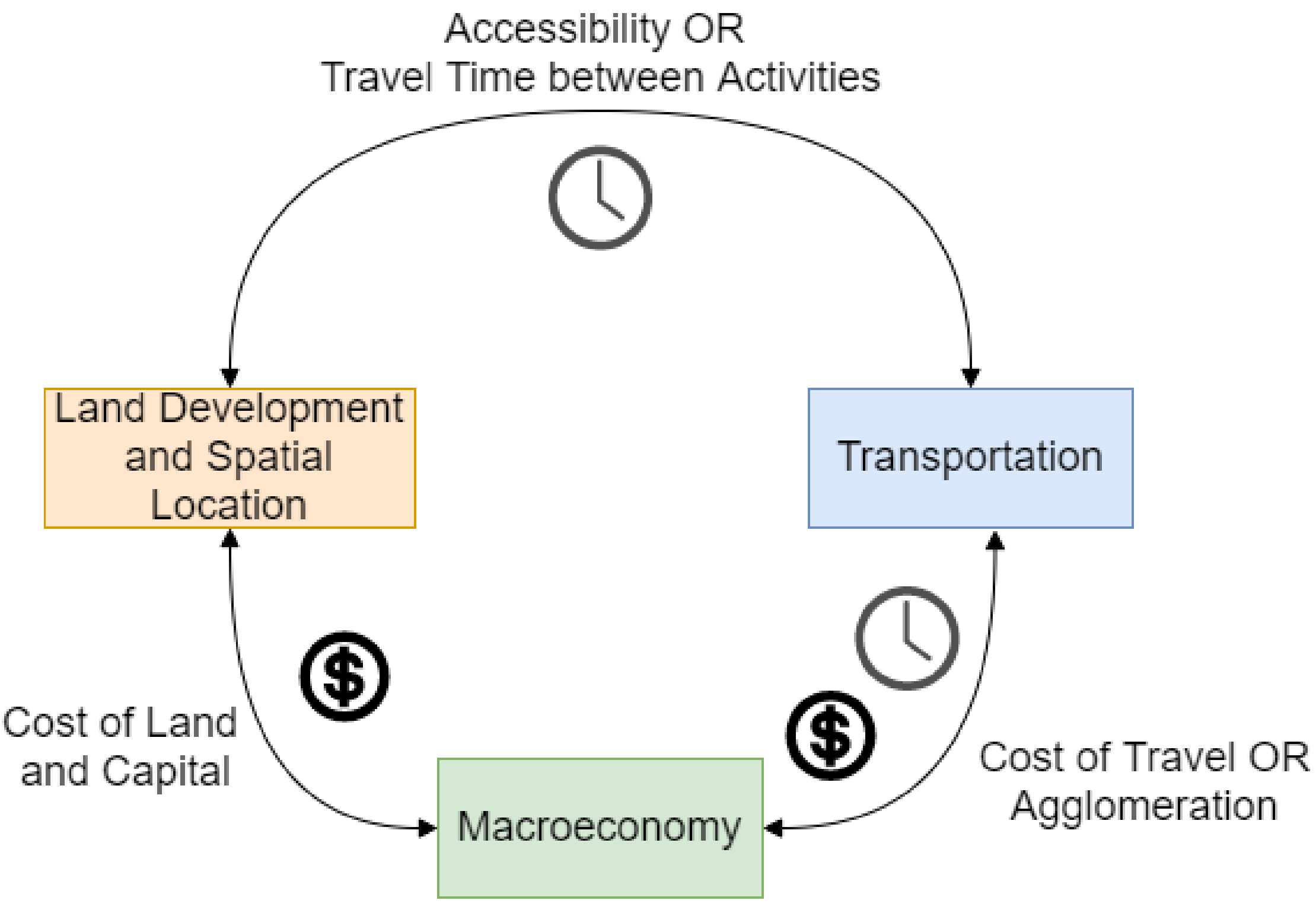
Accessibility OR
Travel Time between Activities
Land Development and Spatial Location
Transportation
Cost of Land and Capital
Macroeconomy
Cost of Travel OR Agglomeration

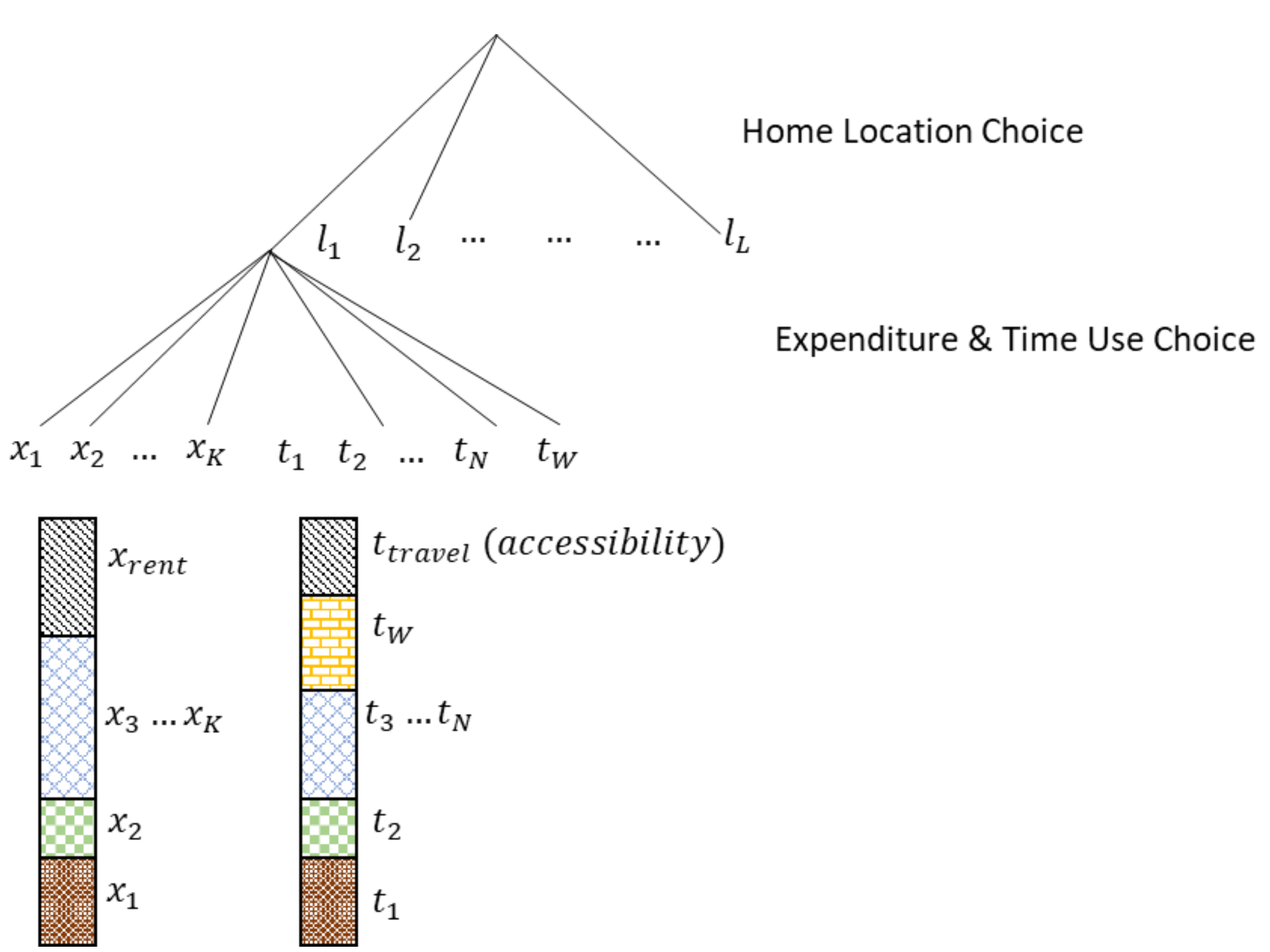
Home Location Choice
$l_1$ $l_2$ ... ... ... $l_L$
Expenditure & Time Use Choice
$x_1$ $x_2$ ... $x_K$ $t_1$ $t_2$ ... $t_N$ $t_W$
$x_{rent}$
$x_3 \dots x_K$
$x_2$
$x_1$
$t_{travel}$ $(accessibility)$
$t_W$
$t_3 \dots t_N$
$t_2$
$t_1$

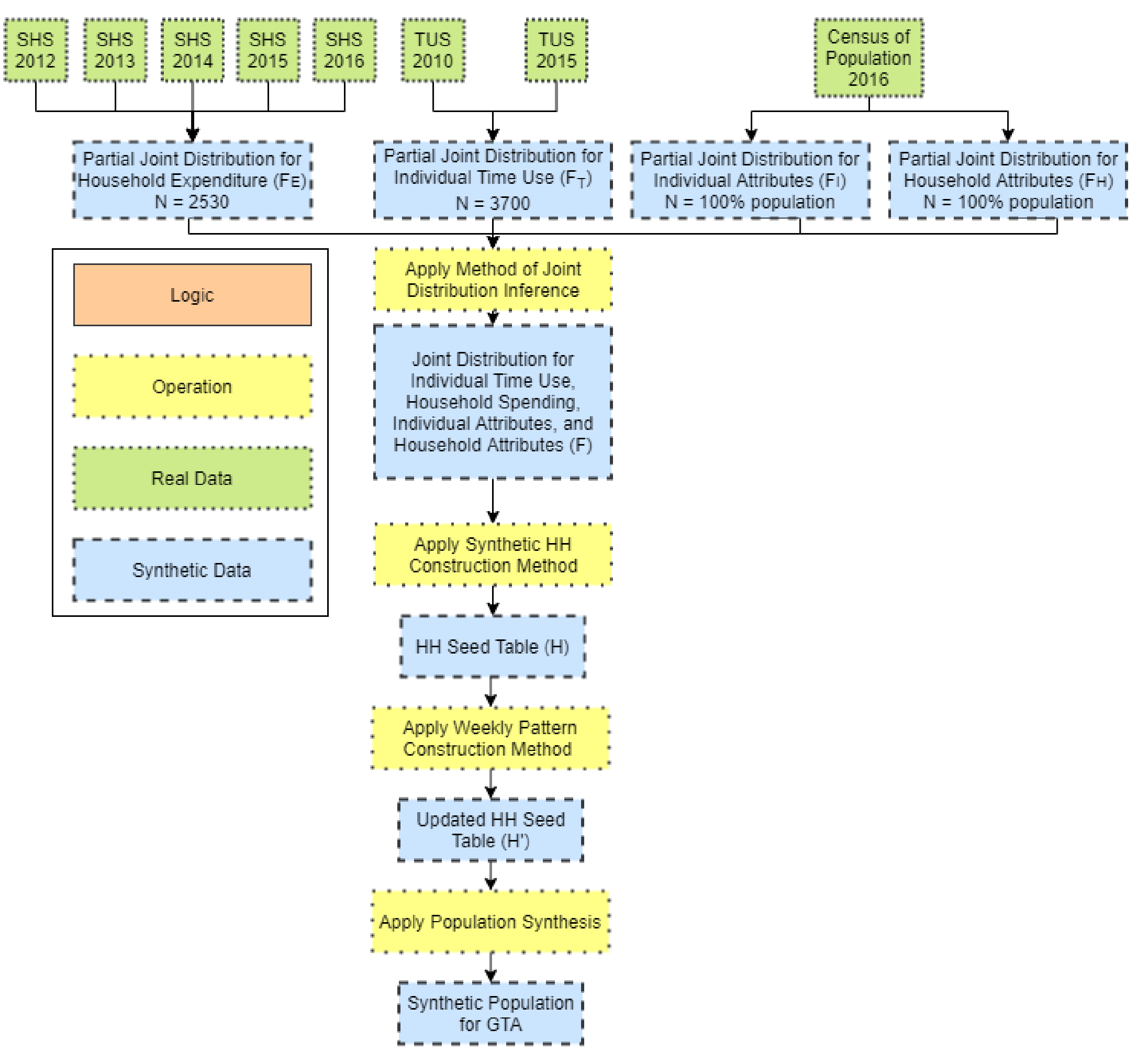

SHS 2012
SHS 2013
SHS 2014
SHS 2015
SHS 2016
TUS 2010
TUS 2015
Census of Population 2016
Partial Joint Distribution for Household Expenditure (FE) N = 2530
Partial Joint Distribution for Individual Time Use (FT) N = 3700
Partial Joint Distribution for Individual Attributes (FI) N = 100% population
Partial Joint Distribution for Household Attributes (FH) N = 100% population
Logic
Operation
Real Data
Synthetic Data
Apply Method of Joint Distribution Inference
Joint Distribution for Individual Time Use, Household Spending, Individual Attributes, and Household Attributes (F)
Apply Synthetic HH Construction Method
HH Seed Table (H)
Apply Weekly Pattern Construction Method
Updated HH Seed Table (H')
Apply Population Synthesis
Synthetic Population for GTA